\documentclass{article}

\usepackage{microtype}
\usepackage{graphicx}
\usepackage{amsmath}
\usepackage{subcaption}
\usepackage{booktabs} % for professional tables
\usepackage{soul} % for highlighting

\usepackage{hyperref}

\usepackage[accepted]{mlsys2021}

\newcommand{\RED}[1]{\textcolor{red}{#1}}
\usepackage{todonotes}
\graphicspath{{./figs/}}
\usepackage{multirow}
\usepackage{multicol} % for \multirow
\usepackage{array}

\mlsystitlerunning{
AnalogNets: Noise-Robust TinyML Models for Analog Compute-in-Memory}

\begin{document}

\twocolumn[
%\mlsystitle{Submission and Formatting Instructions for MLSys 2021}
%\mlsystitle{AnalogNets: Model-HW Co-Design of Noise Robust Deep Neural Networks for Compute-in-Memory TinyML Applications}
\mlsystitle{AnalogNets: ML-HW Co-Design of Noise-robust TinyML Models and Always-On Analog Compute-in-Memory Accelerator}

% unless you've provided the [accepted] option to the mlsys2021


\mlsyssetsymbol{equal}{*}

\begin{mlsysauthorlist}
\mlsysauthor{Chuteng Zhou}{Arm_Boston}
\mlsysauthor{Fernando Garcia Redondo}{Arm_Cam}
\mlsysauthor{Julian B\"{u}chel}{IBM,ETH}
\mlsysauthor{Irem Boybat}{IBM}
\mlsysauthor{Xavier Timoneda Comas}{IBM}
\mlsysauthor{S. R. Nandakumar}{IBM}
\mlsysauthor{Shidhartha Das}{Arm_Cam}
\mlsysauthor{Abu Sebastian}{IBM}
\mlsysauthor{Manuel Le Gallo}{equal,IBM}
\mlsysauthor{Paul N. Whatmough}{equal,Arm_Boston}
\end{mlsysauthorlist}

\mlsysaffiliation{Arm_Boston}{Arm Research, Boston, US}
\mlsysaffiliation{Arm_Cam}{Arm Research, Cambridge, UK}
\mlsysaffiliation{ETH}{ETH Z\"{u}rich, Z\"{u}rich, Switzerland}
\mlsysaffiliation{IBM}{IBM Research, Z\"{u}rich, Switzerland}

\mlsyscorrespondingauthor{Chuteng Zhou}{chu.zhou@arm.com}

\mlsyskeywords{Machine Learning, MLSys}

\vskip 0.3in

\begin{abstract}
%This document provides a basic paper template and submission guidelines.
%Abstracts must be a single paragraph, ideally between 4--6 sentences long.
%Gross violations will trigger corrections at the camera-ready phase.

Always-on TinyML perception tasks in IoT applications require very high energy efficiency.
Analog compute-in-memory (CiM) using non-volatile memory (NVM) promises high efficiency and also provides self-contained on-chip model storage.
%of model weights.
However, analog CiM introduces new practical considerations, 
%model design and training challenges, 
including conductance drift, read/write noise, fixed analog-to-digital (ADC) converter gain, etc.
%noise and non-idealities during inference time, limitations relating to quantization, and so on.
These additional constraints must be addressed to achieve models that can be deployed on analog CiM with acceptable accuracy loss.
This work describes \textit{AnalogNets}: TinyML models for the popular always-on applications of keyword spotting (KWS) and visual wake words (VWW).
The model architectures are specifically designed 
%The model architectures are co-designed 
for analog CiM, 
%to achieve high hardware utilization and model accuracy. 
and we detail a comprehensive training methodology, to retain 
%such that they can better retain their 
accuracy in the face of analog non-idealities, and low-precision data converters at inference time.
%when deploying on analog CiM hardware, especially with low-precision data converters for better energy efficiency.
We also describe AON-CiM, a programmable, minimal-area phase-change memory (PCM) analog CiM accelerator, with a novel layer-serial approach to remove the cost of complex interconnects associated with a fully-pipelined design.
We evaluate the AnalogNets on a calibrated simulator, as well as real hardware,
%to characterize the accuracy after deployment. 
and find that accuracy degradation is limited to 0.8\%/1.2\% after 24 hours of PCM drift (8-bit) for KWS/VWW.
%We also evaluate both models 
AnalogNets running on the 14nm AON-CiM accelerator demonstrate 8.58/4.37 TOPS/W for KWS/VWW workloads using 8-bit activations, respectively, and increasing to
%26.76/12.82 TOPS/W and 
57.39/25.69 TOPS/W 
%at $6$b and 
with $4$-bit activations.
%, efficiencies achieved at the expense of $0.4\%/2.5\%$ and $6.1\%/6.3\% $ accuracy degradation. The accelerator area is $3.2$ mm2}.
\end{abstract}
]



\printAffiliationsAndNotice{\mlsysEqualContribution} % otherwise use the standard text.

\section{Introduction}
\label{sec:intro}

\begin{figure}[ht!]
\begin{center}
\hspace{-0.4cm}
\includegraphics[width=0.45\textwidth]{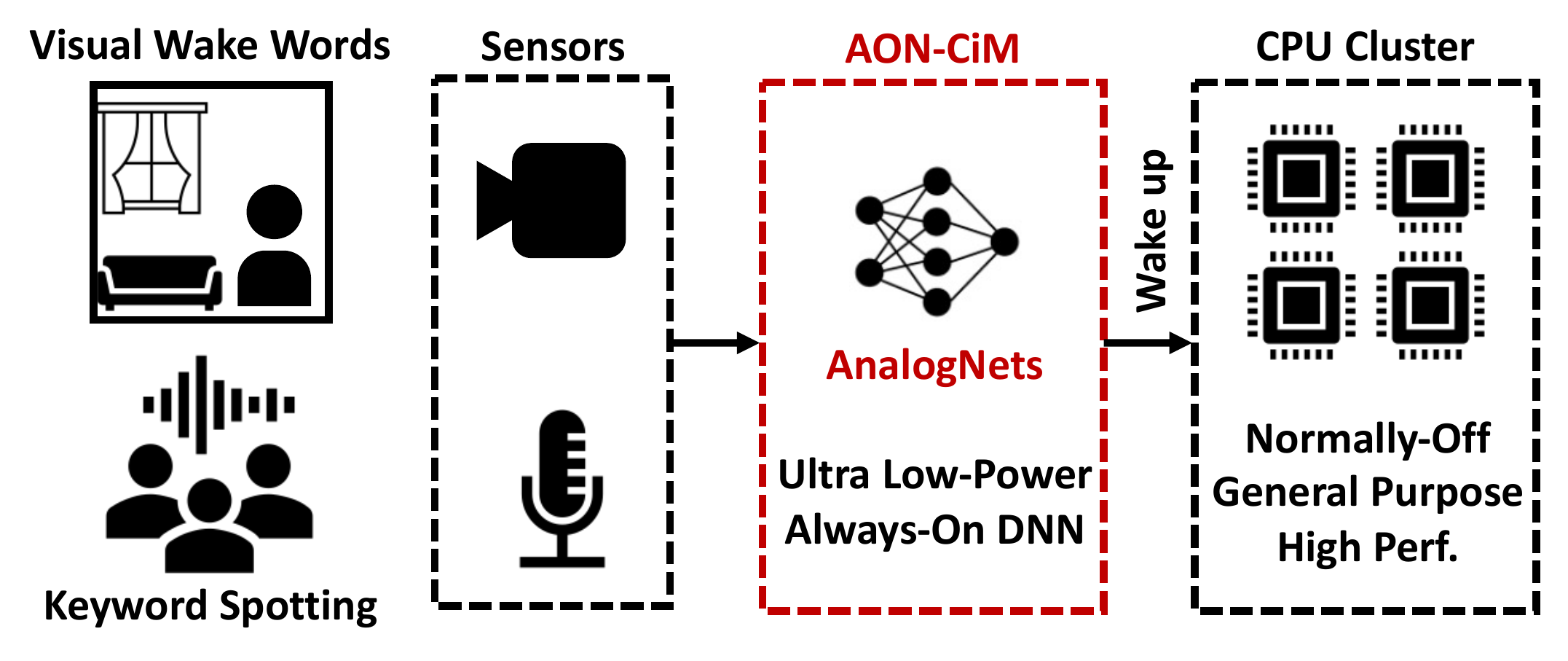}
\end{center}
\vspace{-0.5cm}
\caption{
Overview of always-on (AON) TinyML AnalogNets DNNs on self-contained compute-in-memory (CiM) hardware.
%such as visual wake words (VWW) and keyword spotting (KWS).
}
\label{fig:AON_ML}
\vspace{-0.6cm}
\end{figure}

The origins 
%of so-called 
of deep learning were with high-performance GPU hardware consuming hundreds of Watts of power and costing thousands of dollars.
%Deep neural networks (DNNs) are now ubiquitous for implementing perception tasks 
%on data from various sensor modalities in 
%on internet-of-things (IoT) devices.
%, including audio, image, video and beyond.
However, 
%due to the ubiquity of DNNs, 
there is currently great interest in developing networks that can be deployed for inference tasks on constrained devices that can run autonomously on a few milliwatts of power and cost a few dollars, which is often referred to as \textit{TinyML}.
The applications for TinyML are numerous and varied, but including things like smart hearing aids~\cite{fedorov2020tinylstms}, machinery maintenance monitoring~\cite{banbury2021micronets}, human activity recognition~\cite{kodali_iccd17} and so on. In this paper we focus on two popular always-on perception tasks: visual wake words (VWW)~\cite{chowdhery2019visual} and key-word spotting (KWS)~\cite{warden2018speech}, which turn on the more sophisticated ML functions when a person is in the camera frame or when a keyword is heard (Figure~\ref{fig:AON_ML}).
%shows a schematic of these tasks. 

% TinyML hardware platforms
In many TinyML scenarios, the limitations of a constrained hardware platform introduce very significant challenges for DNN deployment.
Typical microcontroller (MCU) hardware platforms are cheap ($\sim$\$1) and low power ($\sim$10s mW)~\cite{sparse_neurips19}.
However, they often have very limited persistent storage ($\sim$1 MB), memory ($\sim$100s KB) and compute throughput ($\sim$100s MOPS)~\cite{banbury2021micronets}.
This limits the design of the DNN model in terms of: 1) number of parameters, 2) activation footprint, and 3) number of operations (ops) per inference.
These three factors will largely determine the accuracy, latency and energy per inference that can be achieved by a given model.
%on a tinyML task.

% TinyML -> CiM
The high compute and memory demands of DNN inference workloads has recently driven significant research interest in analog compute-in-memory (CiM) hardware platforms~\cite{Zidan2018, Hamdioui2019,memtech_dac19,Y2020sebastianNatNano}. 
The basic idea is to perform computation directly on data stored in on-chip memory, using analog signal processing.
This potentially offers a big gain in energy efficiency compared to digital, due to: 
1) removal of the weight memory read bottleneck, and
2) massive parallelism afforded by a dense array of millions of memory devices performing computation.
%1) huge inherent parallelism owing to removal of the memory read bottleneck, and 2) cheap analog compute primitives based on simple linear circuits.
%from an increase in memory read bandwidth. 
%CiM also offers significant inherent parallelism, with many thousands of multiply-and-accumulate (MAC) operations running in parallel in CiM crossbars in a single execution cycle ($\sim$100 ns).
%on the order of a few hundreds of nanoseconds).
Analog CiM has been demonstrated with a range of underlying memory technologies, including SRAM and non-volatile memory.
Non-volatile memory (NVM) approaches are particularly relevant to TinyML applications, where we need single-chip solutions with all memory on-chip, and with low leakage.
%Relevant NVMs include Flash, phase-change memory (PCM)\RED{Do we need references for PCM etc.?}, resistive random access memory (RRAM) and magnetic random access memory (MRAM); all of which consume negligible static power.
%and are therefore particularly interesting for low power applications. 
%\hl{Similarly, CiM targets the deployment of ML algorithms in constraint systems such as on-body always-ON flexible sensors where digital approaches are inconceivable} \cite{Douthwaite2019}.
A recent phase change memory (PCM) based CiM chip~\cite{Y2021khaddamaljamehVLSI} demonstrated $\>$10 TOPS/W at full utilization.

\begin{figure*}[!t]
\begin{center}
\includegraphics[width=\textwidth]{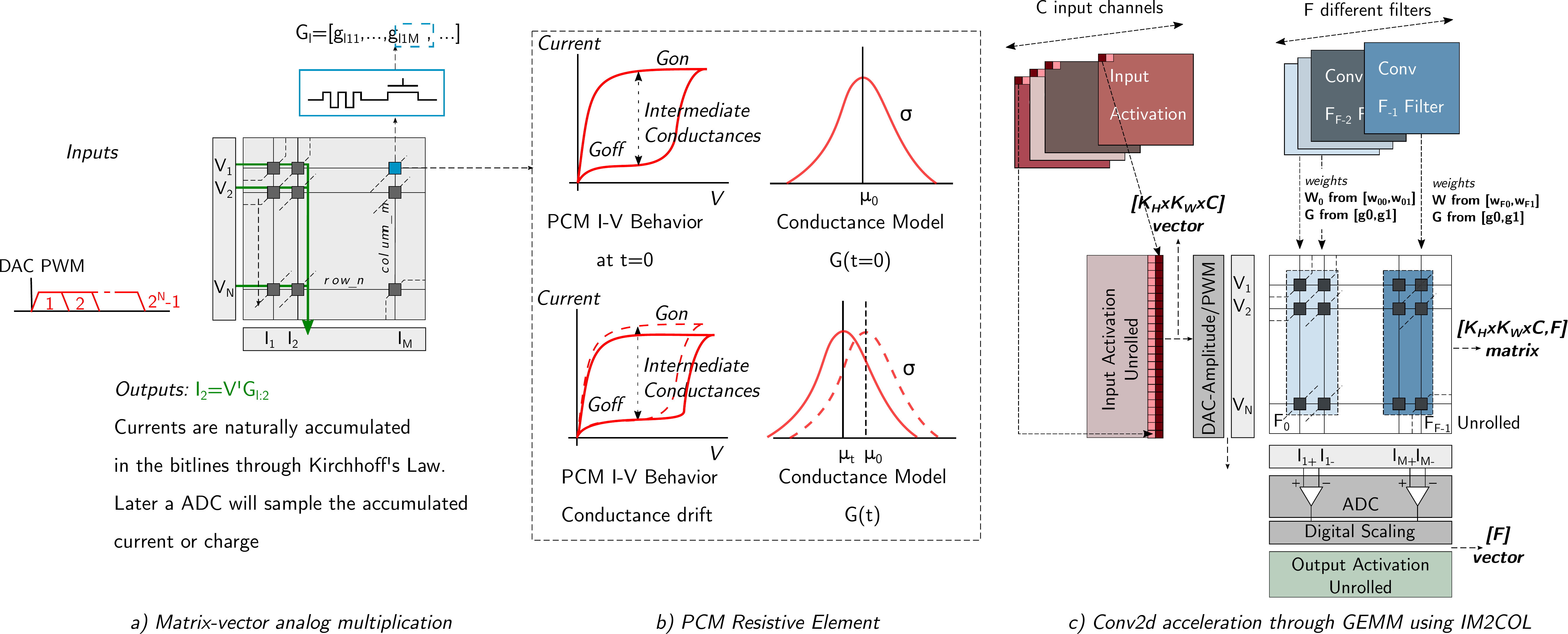}
\end{center}
\vspace{-0.5cm}
\caption{
(a) Matrix-Vector multiplication using analog CiM. 
The weight matrix $\mathbf{W_l}$ is programmed into an array of NVM devices, which provide differential conductances $\mathbf{G_l}$ for analog multiplication, with the input vector encoded as voltage pulses $[V_1, V_N]$. 
The bitlines naturally accumulate the associated charge for each of the currents flowing $[I_1, I_M]$ implicitly performing accumulation.
%in parallel the MAC operations.
(b) The NVM cell conductances, programmed between max/min levels ($G_{ON}-G_{OFF}$), drift over time, introducing a major source of noise in the computation.
%, depending on the characteristics of the NVM technology. 
%In this work we assume that the programmed conductance deviates from the target value following a known distribution, a random Gaussian $\Delta\mathbf{W^l}$ therefore accounting for the inherent imprecision in analog computation.
(c) Deploying a convolution layer, $l$, on analog CiM by flattening the input data and filters into 2D matrices.
%multiplication.
%der to perform a GEMM operation. 
%Digital inputs are translated to the analog domain through DAC modules, while ADC modules convert the analog signals back to the digital domain.
}
\label{fig:xbar}
\end{figure*}

% CiM NN challenges
However, while analog CiM has been motivated as potentially offering orders of magnitude improvements in energy efficiency,
%in comparison to traditional digital accelerators, 
its analog nature introduces a number of new challenges for practical DNN model design and training.
%, which have not been systematically studied to date.
% noise
Firstly, CiM introduces various types of noise in the weight storage, computation, and conversion from digital to analog and back again.
This noise must be accounted for in the model training process to prevent degradation in runtime inference accuracy~\cite{Garcia-Redondo2020, Y2020joshiNatComm}.
%Finally, and affecting both resistive switching and CMOS technologies,
%variability and crossbar-noise caused errors heavily hinder RRAM-CMOS systems
%Even though naturally robust against the noise problems, 
%analog computing blocks approximate the result and are more sensitive to errors.
%This mixed-signal computation noise, together with limited precision components may compromise the signals dynamic range \cite{Garcia-Redondo2020}: On the one hand, the analog noise will depend on the periphery (DACs and ADCs), and the NVM cell static (conductance-programmability) and dynamic (drift, read noise) technology.
%On the other hand, the sizes of the different kernels will determine the amplitudes of electric signals flowing through the system.
%\todo{@Paul, DWC?}
%Different kernel types/sizes involve different SNR, and
%therefore the reliable execution of NN kernels on CiM accelerator requires from the ML-HW co-design.
Secondly, low-bitwidth quantization of models deployed on analog CiM platforms is also challenging, as it is essential to ensure high signal-to-noise (SNR) ratios to mitigate various noise processes.
%It is well known how the operands precision \cite{Krishnamoorthi2018,Liu2019} greatly influence the NN accuracy. 
%For CiM systems not only do these quantization characteristics influence the weights size/accuracy trade-off, but the periphery power, energy, timing and area characteristics \cite{Garcia-Redondo2020}. 
Finally, parameter-efficient TinyML models for CiM can not fully exploit some of the techniques extolled for MCU platforms, such as depthwise separable layers~\cite{mobilenetv1}, which must be expanded into a dense form for CiM.
Therefore, systematic model-hardware co-design is required to deploy efficient, robust tinyML models on analog CiM platforms.
%Therefore both algorithm-hardware should be carefully co-design, tailoring the deployed algorithm for a specific system.

% always-on
%Always-on (AON) functions typically run continuously while the remainder of a SoC is inactive in a low-power sleep mode.
%NVM-based CiM is a promising approach for AON perception for two key reasons: 1) CiM approaches promise high energy efficiency, and 2) on-chip NVM memory stores the model weights without needing access to system memory and with negligible leakage power.
To date, research on CiM focuses on high throughput, fully-parallel hardware, with performance evaluated on either peak numbers (assuming 100\% utilization)~\cite{Y2021khaddamaljamehVLSI}, or on large benchmark models such as ResNet-50~\cite{Y2016shafieeISCA, Y2019ankitASPLOS, Y2021JiaISSCC}, which also achieve high utilization.
%or else merely quoting peak performance assuming 100\% utilization.
%only on 
%%, with an emphasis on TOPS/W metrics 
%\RED{Similarly, the throughput and efficiency reported in many works relates to $100\%$ crossbar utilization ~\cite{aljameh-hermes-vlsi2021}, which does not reflect the performance when accelerating real workloads.}
In contrast, this paper describes end-to-end model-hardware co-design of TinyML DNNs with an always-on PCM CiM accelerator.
%Previous work focuses on fairly large-scale applications with very large die size, presumably targeting data center applications.
%Fully-parallel incurs area overheads
%Focus on big models, such as ResNet etc
%Most previous work focuses on just array-level peak tops/w, but doesn't consider system and model design implications
The main contributions are further summarized below:

\begin{itemize}
    \item 
    %We describe parameter-efficient 
    \textbf{AnalogNets}, compact CiM-optimized TinyML DNNs for keyword spotting
    %~\cite{warden2018speech} 
    and visual wake words tasks.
    %~\cite{chowdhery2019visual}.
    Architecture optimizations for analog noise robustness include removing small bottleneck layers and replacing depthwise layers with regular convolution.
    
    \item 
    \textbf{Training Methodology} for AnalogNets, incorporating noise injection for PCM noise, quantizers for finite DAC/ADC precision, and constrained optimization for fixed ADC gain constraints.
    %challenging hardware constraints and enable deployment on CiM platforms with low-precision data converters.  
    
    \item \textbf{Accuracy Evaluation} of trained AnalogNets on both a calibrated simulator and a real PCM prototype chip.
    The chip measurements show $>$95\% and $>$85\% accuracy for KWS and VWW during a 20-hour experiment. 
    This is the first demonstration of TinyML deployed on analog PCM CiM hardware with competitive accuracy.
    %, previous KWS results achieve 90\% on a simpler KWS task~\cite{guo-rnn-vlsi2019}.
    %Including noise injection, quantization with fixed ADC gain, distillation, etc.
    %\item Evaluate the models on calibrated simulator and real PCM analog CiM.
    
    \item \textbf{AON-CiM}, a minimal area programmable CiM accelerator in 14nm with layer-serialized operation.
    Simulation results show 8.58/4.37 TOPS/W on AnalogNets KWS/VWW with 8-bit activations, up to 57.39/25.69 TOPS/W with 4-bit activations (6.1\%/4.2\% further accuracy degradation after 24 hours).
    %, \RED{XXX}$\timesbaseline MCU platform, when executing the AnalogNets models.
\end{itemize}

\section{Related Work}
\label{sec:related}

%Previous work broadly falls under the following categories.
%: studying the effect of analog computation noise, analysis of noise-injection for DNNs, and use of distillation in model training. 

\paragraph{Models for TinyML}

Mounting interest in TinyML has sparked research on bespoke DNN models specifically targeting highly constrained hardware platforms.  
Early work by \citet{protonn-icml17} and \citet{bonsai-icml17} proposed non-neural ML models to enable inference tasks on MCU hardware platforms with a few KBs of storage.
\citet{sparse_neurips19} demonstrated that CNNs are also feasible on MCUs, by designing a classifier for the CIFAR10-binary task with just KBs of Flash and RAM.
This was achieved using neural architecture search (NAS) to find models that fit the severe model size constraints and still achieve good accuracy.
\citet{fedorov2020tinylstms} demonstrate LSTMs optimized to fit on MCUs for smart hearing-aid applications.
MCUNet~\citep{lin2020mcunet} combines NAS and a custom MCU inference runtime to demonstrate ImageNet models with $>$70\% accuracy on MCUs.
MicroNets~\citep{banbury2021micronets} also used NAS to design state-of-the-art compact models for MCUs.
% employing depthwise separable layers extensively~\citep{mobilenetv1}.
This paper is the first to demonstrate TinyML models for analog CiM hardware.
%, which we develop starting from the SOTA MicroNets models.
%CNN architectures.
%, which we subsequently modify and train to suit analog CiM hardware platforms.

% Stuff from Nic Lane's group?

\paragraph{Computation Noise and Noise-Robust Training} 
In \citet{rekhi2019analog}, noise due to analog computation is modeled as additive parameter noise with zero-mean Gaussian distribution and its variance is a function of the effective number of bits of the output of an analog computation.
Similarly, 
%the authors in 
\citet{Y2020joshiNatComm} also model analog computation using an additive Gaussian on the parameters, with the variance calculated from the observed conductance distributions on their PCM devices. 
%While some studies \citep{ambrogio2018equivalent} have demonstrated \textit{training} as well as inference on analog hardware, training on noisy hardware results in a network with severely degraded accuracy. 
Some previous noise models have even included detailed device-level interactions, such as voltage drop across the analog array \citep{jain2018rx, feinberg2018making}.
%, but are beyond the scope of this paper. 
Training neural networks with random noise \citep{srivastava2014dropout, noh2017regularizing, li2016whiteout, buchel2021} has been demonstrated to be highly effective at reducing overfitting and to increase robustness to adversarial attacks \citep{rakin2018parametric}.
In this work, we employ an additive Gaussian noise model on the weights during training, similar to \citet{rekhi2019analog, Y2020joshiNatComm}. 
Our 
%and present a novel 
noise-aware training method extends previous work to consider low-bitwidth data conversion, leveraging quantization-aware training~\cite{jacob2018quantization}.

\paragraph{Compute in Memory Hardware}
%Various practical analog hardware devices for DNN inference have been demonstrated to date.
%There is also commercial interest in these approaches.
%TODO cite Mythic, lightmatter, lightelligence and any others?
%We divide the various analog DNN inference schemes into electronic and optical.
Analog CiM schemes can be loosely categorized by the type of memory device employed.
Many previous works demonstrate CiM using SRAM
%bitcells
%include %switched-capacitor~\cite{bankman-jssc2018}, 
%charge-based
\cite{Y2019valaviJSSC, biswas2018conv, gonugondla201842pj,Y2021JiaISSCC,lee-sramcim-vlsi2021,tsmc-jssc20}.
%,\citep{bankman2018always, 
SRAM CiM has the significant advantage of availability in cutting-edge CMOS technologies, but is
%However, the typical 6T foundary bitcell is usually not sufficient and a larger cell must be used.
%However, SRAM is 
limited to storing two-levels per bitcell.
%and require multiple cells to represent a multi-bit value.
%which imposes limitations related to binary networks.
A strong alternative is non-volative memory (NVM) \citep{Y2016shafieeISCA,Y2019ankitASPLOS}, such as flash~\citep{merrikh2017high} or memristive devices, such as metal-oxide resistive RAM (ReRAM)~\citep{hu2018memristor, yao2020fully}, and phase-change memory (PCM)~\citep{boybat2018neuromorphic, ambrogio2018equivalent,narayanan-vlsi2021}.
%One of the big advantages of 
Most NVM devices can store multiple levels per cell.
For instance, PCM devices that can compute with an equivalent 8-bit precision ~\cite{Y2018giannopoulosIEDM} have been demonstrated.
%- the achieved bit precision (4-bit for PCM - Le Gallo Nature Electronics 2018, 8-bit in PCM with device level innovations - Giannopoulos 2018 IEDM)
In this paper we employ PCM-based CiM, modeled after~\citet{Y2021khaddamaljamehVLSI}.

%Nvidia DAC paper
%\cite{fixy2019sysml}

\section{Analog CiM Background}
\label{sec:analog}

% \label{sec:problem}
% \begin{figure*}[h]
% \begin{center}
% \includegraphics[width=0.85\textwidth]{xbar.pdf}
% \includegraphics[width=0.8\textwidth]{figs/HW2ML_schematic.pdf}
% \end{center}
% \caption{
% Deploying a neural network layer, $l$, on an analog in-memory crossbar involves first flattening the filters for a given layer into weight matrix $\mathbf{W^l}$, which is then programmed into an array of NVM devices which provide differential conductances $\mathbf{G^l}$ for analog multiplication.  A random Gaussian $\Delta\mathbf{W^l}$ is used to model the inherent imprecision in analog computation.
% }\label{fig:xbar}
% \end{figure*}

% Basic CiM operation
%Analog compute in memory typically uses NVM crossbar arrays to encode the network weights as analog values.
This work focuses on NVM CiM arrays using 
%have been demonstrated using metal-oxide resistive random-access memory (ReRAM)~\citep{hu2018memristor, yao2020fully} and 
phase-change memory (PCM) cells,
%~\citep{, ambrogio2018equivalent, Y2021khaddamaljamehVLSI, Joshi2020}, 
specifically following a recent prototype in silicon by \citet{Y2021khaddamaljamehVLSI}.

\subsection{Analog CiM Operation}
%As depicted in Figure~\ref{fig:xbar}, the 
Matrix-vector operations from DNN inference are performed in parallel inside the CiM crossbar array (Figure~\ref{fig:xbar}).
Weight values are stored in NVM cell array as conductances $\mathbf{G_l}$.
Input activation vectors go through digital-to-analog converters (DACs), which generate pulse-width modulated (PWM) signals $V_N$, applied to the crossbar source lines.
PWM encoding avoids issues with current/voltage non-linearity in the NVM cells.
%The activations are translated into  pulse amplitudes (i.e. DAC amplitude encoding) or pulse durations (i.e. pulse-width modulation or PWM). 
A voltage pulse on a source line results in a momentary current flowing into the bitline from each connected unit cell, proportional to cell conductance $\mathbf{G_l}$. 
All source lines operate in parallel, with the currents from NVM cells summing on the bit lines simultaneously, following Kirchoff's current law. 
Finally, an analog-to-digital converter (ADC) on the end of each bit line integrates the current over the activation pulses, to give a digital pre-activation output.
Outside of the CiM array, bias, batch normalization and activation functions are applied in the digital domain, according to the network architecture.
%further operations such as batch-normalization or activation functions can be applied. 
Note that, a differential NVM cell (two devices) and bitline, is used such that we can represent \textit{signed} weights in $\mathbf{G_l}$.
%, one device for positive and one device for negative. An access device is connected in series to each NVM device to counter sneak path leakage. 

\subsection{Analog Non-Idealities}
%The crossbar architecture automatically computes the addition of the individual dot-products, performing with constant latency as many columns as required, therefore significantly improving the operation throughput, and together with the low-power consumption of the NVM cells, in a very energy efficient manner.
%Our work is based on \citet{aljameh-hermes-vlsi2021}, which demonstrates a 256$\times$256 CiM array, with a cycle time of 130ns, power consumption of 100mW, a peak throughput of 1 TOPS and peak energy efficiency of 10 TOPS/W with 8-bit activations.
%~\citep{Y2021khaddamaljamehVLSI}. The efficiency is expected to increase with larger crossbar sizes. 
%However, the analog nature of both the weight storage and the computation itself limits the precision of the matrix-vector operations, 
%the compute noisy and therefore, the matrix-vector operations can be performed with limited precision. This typically 
%which can result in degraded inference accuracy.
%, which is unacceptable in most applications. 
%For example, a simple two-layer fully-connected network with a baseline accuracy of $91.7\%$ on digital hardware, achieves only $76.7\%$ when implemented on an analog photonic array~\citep{shen2017deep}.
%Therefore, to deploy DNNs on CiM platforms, we must 
%the challenge of imprecise analog hardware motivates us to study and understand \textit{noisy neural networks}, in order to maintain inference accuracy under noisy analog computation. 
%However, despite the clear potential for high energy efficiency using CiM, 
There are various non-idealities in Analog CiM,
%are typically attributed to variations in: 1) manufacturing process, 2) supply voltage and 3) temperature (\textit{PVT variation} collectively), 
all of which result in noise, which 
%These non-idealities 
can degrade inference accuracy.
%, unless accounted for in the DNN architecture and training.
%For further discussion, the reader may find an analysis on how each module  influence the signal-noise ratio throughout the analog system at \cite{Garcia-Redondo2020}.
%In this work, we focus on PCM based CiM approaches, which have been demonstrated extensively~\cite{boybat2018neuromorphic, ambrogio2018equivalent, Joshi2020, Y2021khaddamaljamehVLSI}.
%There is currently no mainstream analog hardware available, so w
%Without loss of generality, we model a noisy machine after a generalized memristive crossbar array, similar to~\citet{shafiee2016isaac}.
Below we discuss the three key practical considerations, which we later address to prevent accuracy loss.
%in DNN architecture and training.
%components in terms of analog noise modeling and some circuit-level considerations.

\subsubsection{NVM Cell Noise and Drift} 
%As described in Figure~\ref{fig:xbar}, 
Resistive PCM cells are each programmed to a value within the minimum and maximum conductance limits (Figure~\ref{fig:xbar}).
%, either continuously or discretely and with different degrees of accuracy depending on the underlying technology (RRAM, PCM, MRAM). 
The programming operation itself has limited precision due to variability and read noise 
%(with a $1/f$ spectral characteristic)
\cite{nandakumar2020precision}.
%Together with this limited and/or inaccurate programming precision, 
In addition, the effective stored value is also dependent on temperature and prone to \textit{drift} over time~\cite{Y2018giannopoulosIEDM,Y2020joshiNatComm}.
While the global component of this variation 
%temperature and drift
%on the matrix-vector operations 
can be compensated to some extent 
by applying digital scaling factors on the ADC outputs \cite{Y2020joshiNatComm},
%However, 
the device-to-device variability 
%of the drift rate and of the conductance temperature dependence will 
still causes noise that cannot be compensated, and degrades the computational precision 
%will inevitably decrease 
over time.

At training time, we generically model this NVM noise as an additive zero-mean \textit{i.i.d.} Gaussian error term on the weights of each layer $\Delta \mathbf{W}_l\sim\displaystyle \mathcal{N} (0 , \sigma_{N,l}^2 \mathbf{I})$, similar to~\citet{Y2020joshiNatComm}. 
%This noise model, described more concretely in Section~\ref{sec:models}, is similar to that used by~\citet{Joshi2020} during training. 
Then, at test time, we employ a detailed statistical PCM CiM model that accurately implements programming noise, 1/f noise and conductance drift, calibrated on the characterization of doped-Ge$_2$Sb$_2$Te$_5$ (d-GST) mushroom-type PCM from a million device array integrated in 90nm CMOS technology \cite{Nandakumar2019}.

\subsubsection{DAC/ADC precision}
%and Digital-to-Analog Data Converters. Precision and noise.}
%As for digital systems, the precision of the operands involved in the analog GEMM computation determines its accuracy. But it is well known that for analog accelerators,  
%In analog CiM accelerators, 
The PWM DACs convert the input digital data to a series of unit voltage pulses on the source lines.
%duration through a Pulse-Width-Modulation (PWM) circuit --or interchangeably a train of pulses--, 
Meanwhile, the ADCs sample the accumulated charge on the bitlines~\cite{yao2020fully,Y2021khaddamaljamehVLSI,narayanan-vlsi2021}. 
The DAC can be therefore abstracted as a quantizer on the input activations, assuming a given effective number of bits (ENOB). The DAC quantization ranges are set in digital, and therefore can be arbitrarily tuned per layer to properly address the variable dynamic range of the signals being converted. However, they must be kept constant throughout inference in order to avoid extra dynamic scaling circuits. Similarly, the ADC can be modeled as output activation ENOB quantization. However, the ADC quantization range cannot be arbitrarily tuned because its analog gain is usually fixed during a calibrating step.

DAC and ADC ENOB have a strong impact on the throughput and energy efficiency CiM approaches~\cite{Li2017e, Ni2017}.
However, quantizing DNNs for CiM is challenging, especially below 8-bits, so the PPA advantage must be balanced against the model accuracy impact.

\subsubsection{ADC Gain}
\label{sec:analog:adc_gain}
The last consideration is the ADC gain, the ADC analog gain stage must be calibrated to achieve a given conversion of bit-line charge to digital values \cite{Y2021khaddamaljamehVLSI}.
%require a calibration stage and stable references, which ideally would remain the same for the whole model~\cite{Garcia-Redondo2020}. 
Abstracted at the ML algorithm level
%, and as described in detail in 
(Section~\ref{sec:models}), this calibration can be represented as a scaling on the output activations prior to quantization. 
Because the time-consuming calibration must be performed for each ADC individually, it is not feasible to recalibrate the ADCs at inference time to achieve optimal quantization ranges per layer.
%for all signals being converted. 
%Moreover, this calibration using analog circuits cannot be expected to achieve floating-point precision tuning of the ADC ranges per layer. 
Therefore, in this work, we require identical ADC gain across all layers.
%and cannot be dynamically tuned during inference. 
%Following that idea, to reduce the module to model calibration requirements, extra scaling stages can be applied to the input and output activations in the digital domain \cite{Y2021khaddamaljamehVLSI}.
%\todo{what does the previous sentence mean?}
%More importantly, a time-dependent variable scaling can be applied to correctly account for conductance drift \cite{Giannopoulos2019,Joshi2020}, compensating for the undesired variations on the conductance value.
%The results presented in Section~\ref{sec:results} follow \cite{Joshi2020} approach, proving the validity of the method.
%{\color{blue} \bf @manuel please can you review this paragraph}

% If added, should be at the end, after the analog noise and non-idealities are described
\if0
The question of how to effectively learn and compute with a noisy machine is a long-standing problem of interest in machine learning and computer science~\citep{stevenson1990sensitivity,von1956probabilistic}. 
In this paper, we study noisy neural networks to understand their inference performance.
We also demonstrate how to train a neural network with distillation and noise injection to make it more resilient to computation noise, enabling higher inference accuracy for models deployed on analog hardware. 
We present empirical results that demonstrate state-of-the-art noise tolerance on multiple datasets, including ImageNet.
\fi

\begin{figure}[t!]
\centering
%\includegraphics[width=0.22\textwidth]{figs/depwthwise_pcm.png}
%\includegraphics[width=0.22\textwidth]{figs/bottleneck_schematic.png}
%\hspace{-0.3cm}
\hspace{0.3cm}
\includegraphics[width=0.45\textwidth]{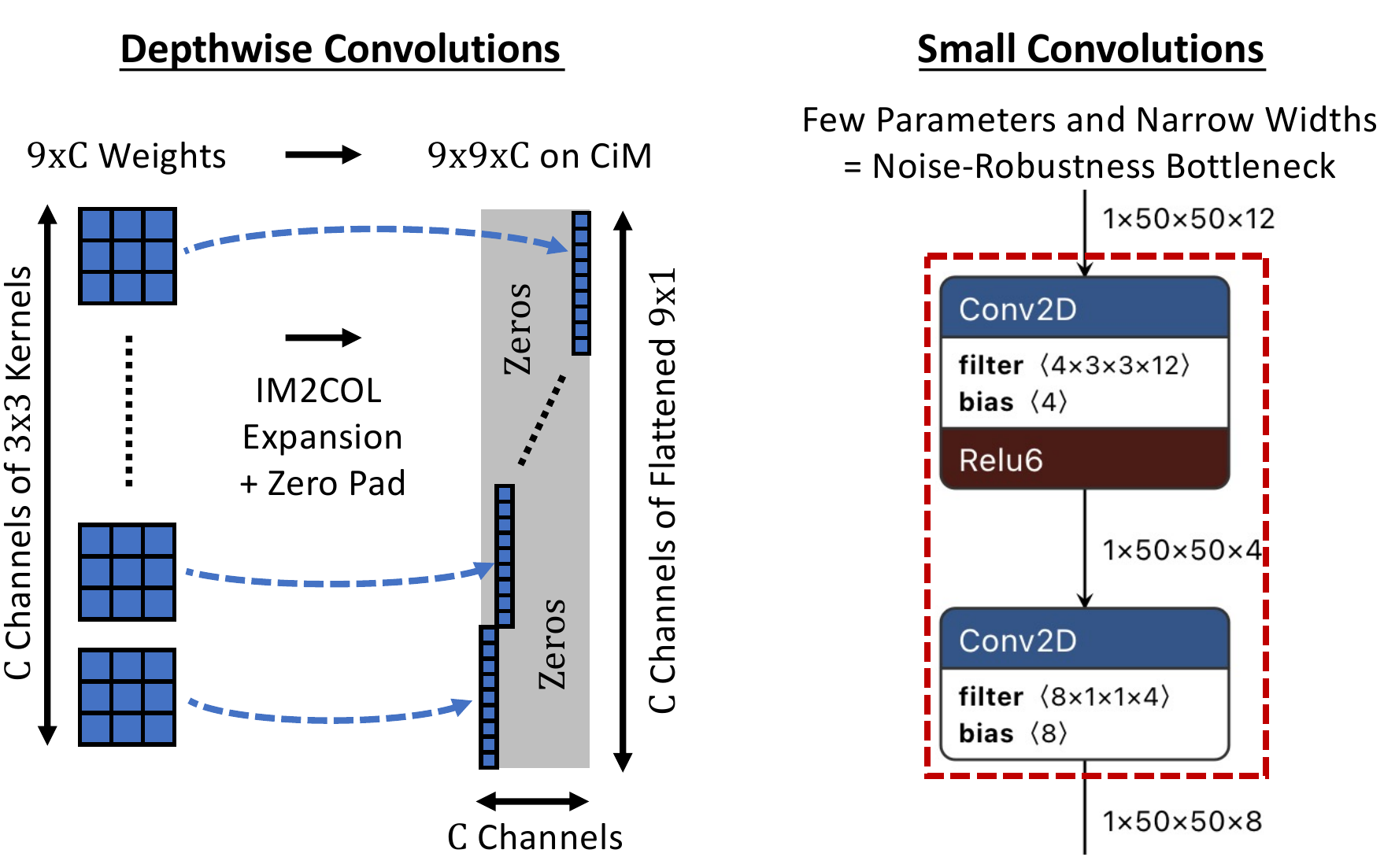}
\vspace{-0.4cm}
\caption{Analog CiM model design: 
(Left) Depthwise convolutions are not suitable for CiM, since: 
1) mapping to CiM incurs IM2COL expansion with zero padding,
%, which leads to two drawbacks: 
increasing the effective storage space by $\sim$9$\times$, and 
2) the signal in each column becomes small, such that analog noise degrades accuracy.
%has an exaggerated impact on accuracy. 
%For these reasons, 
%depthwise convolutions are explicitly avoided in 
Therefore, the AnalogNets architectures instead use
%, in favor of 
simple 3$\times$3 kernels. 
(Right) Small, narrow layers (in the VWW model) are noise-robustness bottlenecks and are therefore avoided in AnalogNets.
%identified 
%as they are robustness bottlenecks.
}
\label{fig:depthwise}
\end{figure}

\section{AnalogNets: TinyML for CiM}
\label{sec:models}

In this section, we build on previous TinyML models to meet the challenges of deploying DNNs on analog CiM hardware.
We use the keyword spotting (KWS) and visual wake words (VWW) models from MicroNets~\cite{banbury2021micronets}. 
These state-of-the-art models are found by Neural Architecture Search (NAS) with constraints on model size, working memory and number of operations, 
%constraints, and are designed 
to run efficiently on MCUs.
%commodity micro-controllers with as little as 0.5MB of flash and 128KB of SRAM. 
We start with the smallest MicroNets models, more specifically MicroNet-KWS-S for KWS and MicroNet-VWW-2 (100$\times$100 resolution) for VWW.

\subsection{Model Architectures}

\paragraph{Depthwise Convolutions Are Not Suitable for CiM}
Most state-of-the-art efficient DNNs 
%deep learning models 
feature depthwise convolution layers, due to their superior parameter efficiency. MicroNets models are no exception. 
Unfortunately, there are two issues with depthwise convolution
%is deployed 
on an analog CiM accelerator.
%, two major challenges are encountered. 
First, 
%the mapping of 
depthwise layers are very inefficient when mapped onto a CiM array, as they must be expanded into a dense form with a non-zero diagonal,
%because a depthwise layer has 
%to the diagonal non-zero mapping which results 
resulting in low hardware utilization (Figure \ref{fig:depthwise}, left). 
For example, when mapped to CiM, the second 3$\times$3 depthwise convolution layer in MicroNet-KWS-S
%serves to illustrate the drawbacks of deploying DWC layer in NVM crossbars. 
%Regarding area requirements, being a layer with  
%has 112 input channels, and the 
has local array utilization as low as $1/112\approx0.9\%$. 
% To improve the area utilization, different depth-wise convolutions can share the same source lines bit are placed at different bit lines, however at the expense of performance and energy \cite{Y2021OttaviAICAS}. --Figure~\ref{fig:depth_wise_kws}.
Second, the unused cells can also contribute additional noise into the bitline, which degrades the SNR (see remarks on small layers shortly).
%The second challenge comes from the accumulation of noise from the unused cells on the same bit line.
%All non-used NVM area will be stressed by the incoming input voltages, and contribute with non-negligible noise levels degrading the bit line SNR.
%Finally, it is to be noted that due to the NVM crossbar inefficient utilization, $99.1\%$ of the crossbar power is wasted.

Simulation of MicroNet-KWS-S on the PCM CiM simulator reveals unsatisfactory accuracy performance. 
The drop in accuracy is almost $10\%$ after a year of drift when employing 8-bit ADCs (see Appendix), compared to $<2\%$ for the model we present in this paper. 
As a general design principle, 
%argue that  ML models
DNNs designed for CiM should avoid depthwise convolutions.
%The way we implement the design principle 
Therefore, in this paper we replace all depth-wise separable convolution blocks by regular 3$\times$3 convolutions. For MicroNet-VWW-2 which is based on a MobileNetV2 backbone, the inverted bottleneck MBConv block thus becomes a fused-MBConv block~\cite{tan2104efficientnetv2}. 
Although this change avoids the problems with depthwise convolution, it also introduces more parameters into the model. 
Therefore, to ensure the two models still fit on the same size CiM array (see Section~\ref{sec:hw}), we have to 
%also make the design decision to 
remove the last parameter-heavy layer of 196 output channels from the KWS model.

\paragraph{Small Layers Are Bottlenecks} 
Efficient models may have narrow layers in the beginning of the network to keep the memory footprint low. 
But, we further found that these very narrow layers are bottlenecks that limit accuracy in the presence of noise.
%in terms of accuracy. 
Narrow layers have few parameters and therefore are more sensitive to noise
%, the hardware non-ideality has a bigger impact on those layers 
(in terms of variance). 
Layers with a small number of parameters 
%also means that it 
may also not have enough redundancy to gain noise robustness during training. 
Theoretical work on information decay in noisy neural networks also supports avoiding narrow bottleneck layers~\cite{ZhouISIT2021}.
In our VWW model, we have identified two early layers that are bottlenecks, shown in Figure~\ref{fig:depthwise} (right), which 
%Deploying them on the CiM simulator 
result in degraded accuracy performance when tested on the CiM simulator. 
Therefore, to maintain accuracy, we remove these layers completely, which we find has little impact on model accuracy, model size and working memory.
%requirement. 
Table~\ref{tab:ablation} shows the simulated accuracy of these two different model design choices (last and second last rows).
%, to validate our solution. 

The two models developed
%following the observations above to co-design for analog CiM hardware, 
are subsequently referred to as 
%For naming convenience, the two models that have gone through these HW-informed transformations are called 
AnalogNet-KWS and AnalogNet-VWW. Their exact architectures can be found in the Appendix.

\subsection{Model Training Methodology}\label{sec:train_methodology}
\begin{figure}[t!]
\begin{center}
\includegraphics[width=0.5\textwidth]{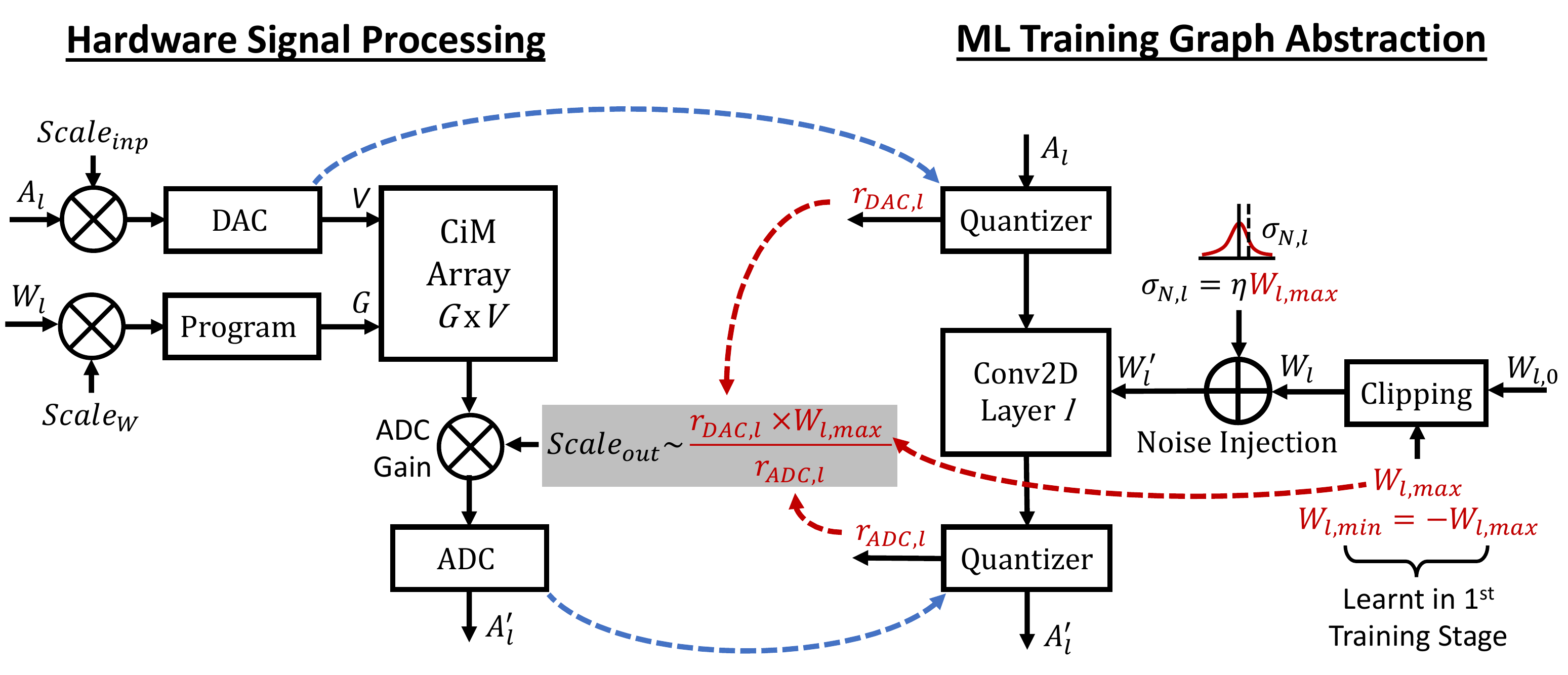}
\end{center}
\vspace{-0.4cm}
\caption{
Analog CiM hardware constraints (left) mapped to our HW-aware training procedure that incorporates noise injection, as well as DAC and ADC constraints during end-to-end training.}
\label{fig:HW2ML}
\end{figure}

A CiM-specific training methodology is essential for these models.
%It is important to craft a HW-specific training flow for these models. 
As we will show in Section~\ref{sec:results}, conventional TinyML models are not robust to analog CiM non-idealities, especially at low DAC/ADC bit precision.
%for the ADC/DACs. 
As discussed in Section~\ref{sec:analog}, there are many different aspects of analog CiM hardware that must be properly accounted for during training.
%taken into consideration. 
%It is important to expose model training to these non-idealities and constraints in a practical way. 
However, it is not practial to incorporate every detail of the hardware in the training process, so we develop a set of robust abstractions.
%Having every detail of the analog CiM hardware in the ML model training flow is impractical and 
For example, we simplify modeling of PCM noise (i.e., programming, temperature and drift effects) using Gaussian noise injection in weights.
%to model the errors in programming the network weights in conductance and its temporal fluctuations during inference (read/write errors). 
\citet{Y2020joshiNatComm} described this approach in detail and demonstrated accuracy improvement on ResNet models measured on real PCM-based prototype hardware. 
Our work further develops this simple approach to train TinyML models 
%for edge applications that are 
adapted for inference on CiM with low-precision ADC/DACs. 
Figure~\ref{fig:HW2ML} illustrates the process which abstracts the HW constraints and non-idealities to components in the ML model training graph. We next describe each component.

\paragraph{Noise Injection} At each forward pass, the weight noise in layer $l$ is drawn from an \textit{i.i.d.} Gaussian distribution $\displaystyle \mathcal{N} ( \mathbf{0}, \sigma_{N,l}^2 \mathbf{I})$ and  added to the weights $\mathbf{W}_l$. The noise level is referenced to the maximum weight value in layer $l$ 
\begin{equation}\label{eqn:eta}
    \sigma_{N,l} = \eta W_{l,\mathrm{max}},
\end{equation}
where $\eta$ is a coefficient characterizing the noise power; a higher $\eta$ implies a reduced Signal-to-Noise Ratio (SNR). From a PCM CiM HW point of view, $\eta$ can be viewed as the ratio between combined average conductance noise standard deviation and the maximal conductance range on the PCM array~\cite{Y2020joshiNatComm}. 
% Some work~\cite{Zhou2020a} in the literature explored injecting non-\textit{i.i.d.} noise into weights, in this paper we only present \textit{i.i.d.} noise. 
The maximal weight value $W_{l,\mathrm{max}}$ obtained from regular neural network training is on the very tail of the distribution and highly susceptible to outlier influence. 
Therefore weight clipping is required to control the value of $W_{l,\mathrm{max}}$. We have 
\begin{equation}
    \mathbf{W}_l = \mathrm{clip}(\mathbf{W}_{l,0};W_{l,\mathrm{min}}, W_{l,\mathrm{max}}).
\end{equation}
%There are different ways to choose the clipping ranges.
\citet{Y2020joshiNatComm} compute clipping ranges dynamically, based on statistics of $\mathbf{W}_{l,0}$. Our approach is to use static clipping ranges during noise injection training, which we found is good for training stability. More details on calculation of clipping ranges
%The way these static clipping ranges are computed for each layer 
can be found later in this section. 

In terms of gradient computation, we treat the entire clipping and noise injection operation with a straight-through estimator (STE)~\cite{bengio2013estimating}. The gradients are computed with clipped and noise-perturbed weights, and are then applied to $\mathbf{W}_{l,0}$.

\paragraph{DAC/ADC Constraints}
\label{sec:models:adc_gain}
In Section \ref{sec:analog} we described the role of DACs and ADCs in an analog CiM accelerator.
%and how we model them in this paper. 
From an ML training abstraction,
%point of view, 
we model the DACs and ADCs with simple effective number of bits (ENOB) quantizers. 
Quantization, whether for model weights or activations, is becoming ubiquitous in ML model training and there are many tools available. 
In the ML training graph (Figure \ref{fig:HW2ML}), we model DACs and ADCs as separate quantization nodes. 
The input activation tensor is quantized by the DAC coming into the operator and the output pre-activation tensor is quantized by the ADC. 
Hardware constraints dictate that the ranges of these quantizers are symmetric, and we set the ENOB of the DAC to be 1-bit more than the ADC to accommodate the positivity of input activations after a ReLU activation function:
\begin{equation}
    b_{\mathrm{DAC}} = b_{\mathrm{ADC}} + 1.
\end{equation}
Setting the ranges for these quantizers, or equivalently choosing the scaling/gain factors for DAC/ADCs, is not easy especially because of the varying dynamic ranges between the signals on different filters and layers for a deep neural network with many layers.
Furthermore, for low-precision DAC/ADCs, sub-optimal choices of these quantities leads to accuracy degradation. 
% \RED{An initial solution to this problem was proposed in \cite{Garcia-Redondo2020}, where a custom differentiable training method globally accounting for the dynamic ranges imposes scaling/boundary constraints for the different signals.
% However, due to the sensitivity to the analog noise, this method can only be applied for extremely constraint and small neural networks, and therefore its applicability to TinyML is very limited.}
% \RED{To address this problem and find a solution to larger networks  }
To address this problem, we adopt a fully differentiable approach to learn the ranges using gradient descent. Following a similar approach to~\citet{jain2019trained}, the quantization function can be written as 
\begin{equation}
    q(x;b) = \mathrm{round}_\mathrm{STE}\left(\frac{\mathrm{clip}(x;-r_\mathrm{max},r_\mathrm{max})}{r_\mathrm{max}/(2^{b-1}-1)}\right).
\end{equation}
The rounding operation rounds the number to its nearest integer and its gradient is calculated using the straight-through estimator (STE)~\cite{bengio2013estimating}. 
This quantization function is differentiable with respect to both $x$ and $r_\mathrm{max}$.

Figure \ref{fig:HW2ML} also describes the ADC gain constraint discussed in Section~\ref{sec:analog}. 
The analog signal coming from the bitlines in the CiM array need to be scaled in the analog domain before being converted by the ADC. 
This analog scaling represents the ADC gain and is set at calibration time, and therefore it should be considered a constant across different layers. 
The scaling factor is related to the ranges of quantizers and maximum weight values to make sure that the ranges of ADCs and DACs are fully utilized. 
This constraint can be expressed as 
\begin{equation}
    S \equiv \frac{r_{\mathrm{DAC},l}W_{l,\mathrm{max}}}{r_{\mathrm{ADC},l}}, \forall l. 
\end{equation}
$r_{\mathrm{DAC},l}$ and $r_{\mathrm{ADC},l}$ are respectively the maximum ranges of quantizers representing the DAC and the ADC in layer $l$. 
$W_{l,\mathrm{max}}$ is related to the noise injected into model weights by Equation \ref{eqn:eta}. 
To eliminate one moving piece, we fix the value of $W_{l,\mathrm{max}}$ during a two-stage training process.

% \begin{figure}[t]
% \centering
% \hspace{1.5cm}
% \includegraphics[width=0.35\textwidth]{figs/cim-accelerator.pdf}
% \caption{AON-CiM Accelerator...}
% \label{fig:accelerator}
% \end{figure}
\begin{figure}[t]
\centering
\hspace*{0.1cm}
\includegraphics[width=\columnwidth]{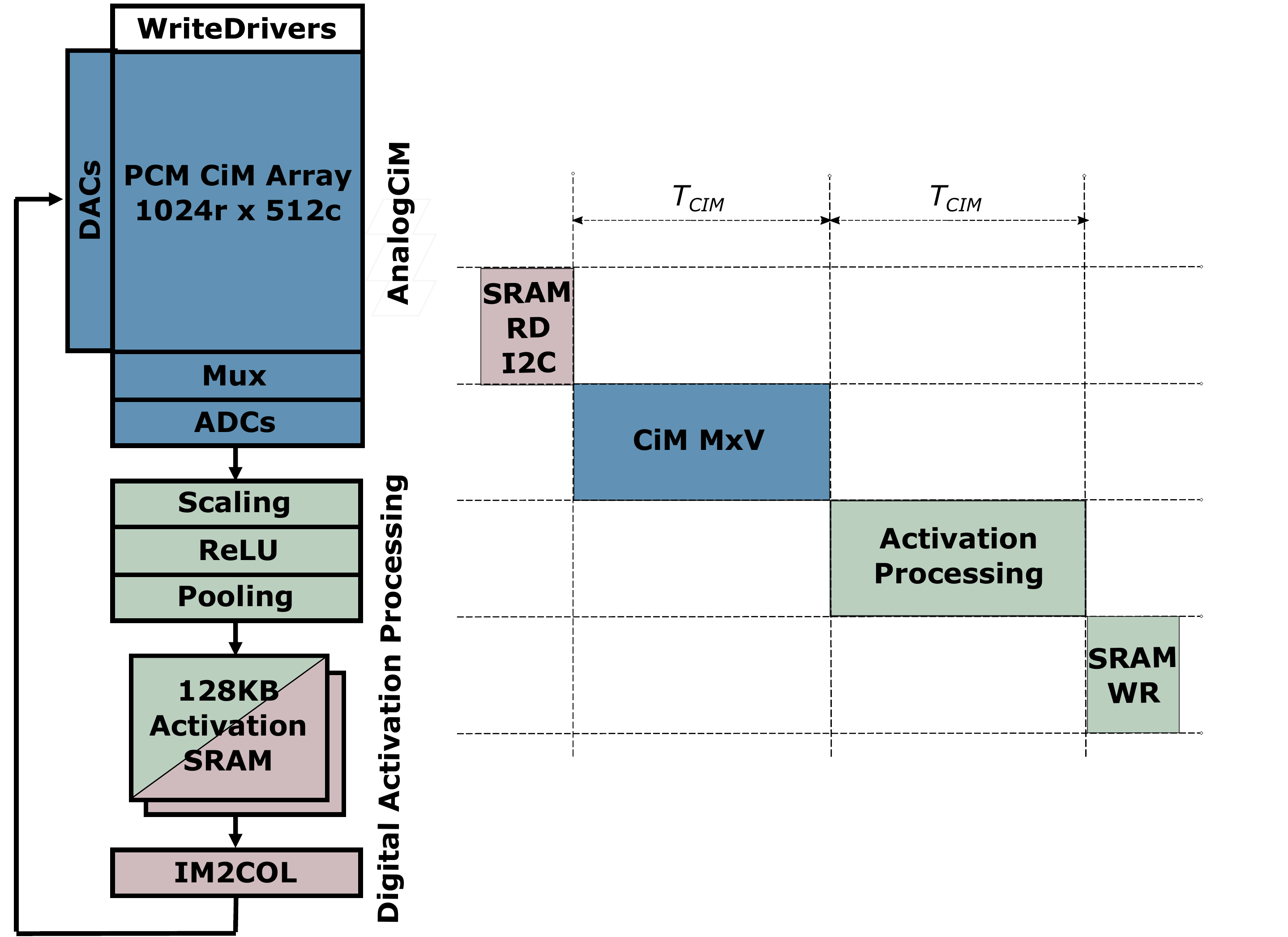}
\vspace{-0.6cm}
\caption{AON-CiM accelerator architecture and operation. 
Following \citet{Y2021khaddamaljamehVLSI}, each ADC quantized value requires a floating point activation/scaling operation pipelined to adjust to the SRAM read/writes and crossbar operation.
}
% \caption{AON-CiM accelerator architecture and operation. 
% \RED{Following \citet{aljameh-hermes-vlsi2021}, each ADC quantized value requires two floating point scaling operations. Depending on the number of FP units, the scaling time $T_{FP}$ varies. An adjusted pipelined process would match the CiM latency so $T_{CIM}=T_{FP}$. Over or under adjusted pipelines would incur into \emph{Idle} states where either the CiM or the FP modules would be stalled.}
% }
\label{fig:accelerator}
\end{figure}

In the first training stage, we train a model without any quantization nodes or noise injection and only with weight clipping in the range $[-2\sigma_{\mathbf{W}_{l,0}},2\sigma_{\mathbf{W}_{l,0}}]$, where $\sigma_{\mathbf{W}_{l,0}}$ is the standard deviation of weights calculated with the unclipped weights $\mathbf{W}_{l,0}$. 
$\sigma_{\mathbf{W}_{l,0}}$ is updated every 10 time steps until the model is trained to convergence.

In the second training stage, we initialize the model with the trained weights from the first stage and compute the clipping ranges as $W_{l,\mathrm{min}} = -2\sigma_{\mathbf{W}_{l,0}}$, $W_{l,\mathrm{max}} = 2\sigma_{\mathbf{W}_{l,0}}$. 
$W_{l,\mathrm{min}}$ and  $W_{l,\mathrm{max}}$ are then fixed throughout the training. We add noise injection and the quantization nodes representing the ADCs and DACs. 
We treat $S$ and $r_{\mathrm{ADC},l}$ as independent trainable parameters and initialize them at $1$. Therefore  $r_{\mathrm{DAC},l}=r_{\mathrm{ADC},l}|S|/W_{l,\mathrm{max}}$ and we take the absolute value of $S$ here to make sure that the ranges are all positive since $S$ may become negative during gradient descent. 
The gradient of $S$ can be computed as 
\begin{equation}
    \frac{\partial \mathcal{L}}{\partial S} =
    \sum_{l} \frac{\partial \mathcal{L}}{\partial r_{\mathrm{DAC},l}}
    \frac{r_{\mathrm{ADC},l}}{W_{l,\mathrm{max}}}\frac{d|S|}{dS}, 
\end{equation}
where $d|S|/dS$ denotes the subgradient for the absolute value function. The gradients of $r_{\mathrm{ADC},l}$ are also slightly altered since there is now an extra differentiable path through $r_{\mathrm{DAC},l}$. There are no gradients for $W_{l,\mathrm{max}}$ as they are constant. Modern deep learning frameworks support automatic differentiation and can handle these gradient calculations seamlessly. We use TensorFlow to train our models.

% \begin{itemize}
%     \item Previous work has demonstrated larger models, e.g. ResNet-Xx trained to accommodate analog imperfections.
%     \item Here, we consider the same for edge / TinyML applications.
%     \item Design of compact models; challenges with depth-wise convolution.
%     \item Training of models for noise tolerance (with/without depthwise conv).
%     \item ADC gain
%     \item Low bitwidth
%     \item distillation
%     \item show models for 3 datasets, with measured accuracy on hardware
% \end{itemize}

% \begin{figure}[t]
% \begin{center}
% \includegraphics[width=\linewidth]{schematic_new.png}
% \end{center}
% \caption{Combining noise injection and distillation.} \label{fig:schematic}
% \end{figure}

\section{AON-CiM Accelerator}
\label{sec:hw}

Always-on (AON) accelerators operate alone while the remaining blocks in an SoC are unused and powered-down.
%, such that only the AON functions are running.
Therefore, an AON accelerator must be self-contained, including all required memory.
Below, we describe the fully-programmable AON-CiM accelerator used to evaluate the AnalogNets models.
%, its architecture, and design characteristics.

% The trim coordinates are: W S E N
\begin{figure}[t]
    \centering
%   \begin{subfigure}[b]{0.49\columnwidth}
%     \hspace{-0.5cm}
%     \includegraphics[width=\linewidth]{figs/baseline/analognet-kws_big/mapping_m_0_r_1024_c_512.pdf}
%     \vspace{-0.8cm}
%     \caption{AnalogNets-KWS}
%     \label{fig:mapping:kws}
%   \end{subfigure}
%   %\hfill %%
%   \begin{subfigure}[b]{0.49\columnwidth}
%     \hspace{-0.4cm}
%     \includegraphics[width=\linewidth]{figs/baseline/analognet-vww_big/mapping_m_0_r_1024_c_512.pdf}
%     \vspace{-0.8cm}
%     \caption{AnalogNets-VWW}
%     \label{fig:mapping:vww}
%   \end{subfigure}
    \hspace{-0.2cm}
    \includegraphics[width=\linewidth]{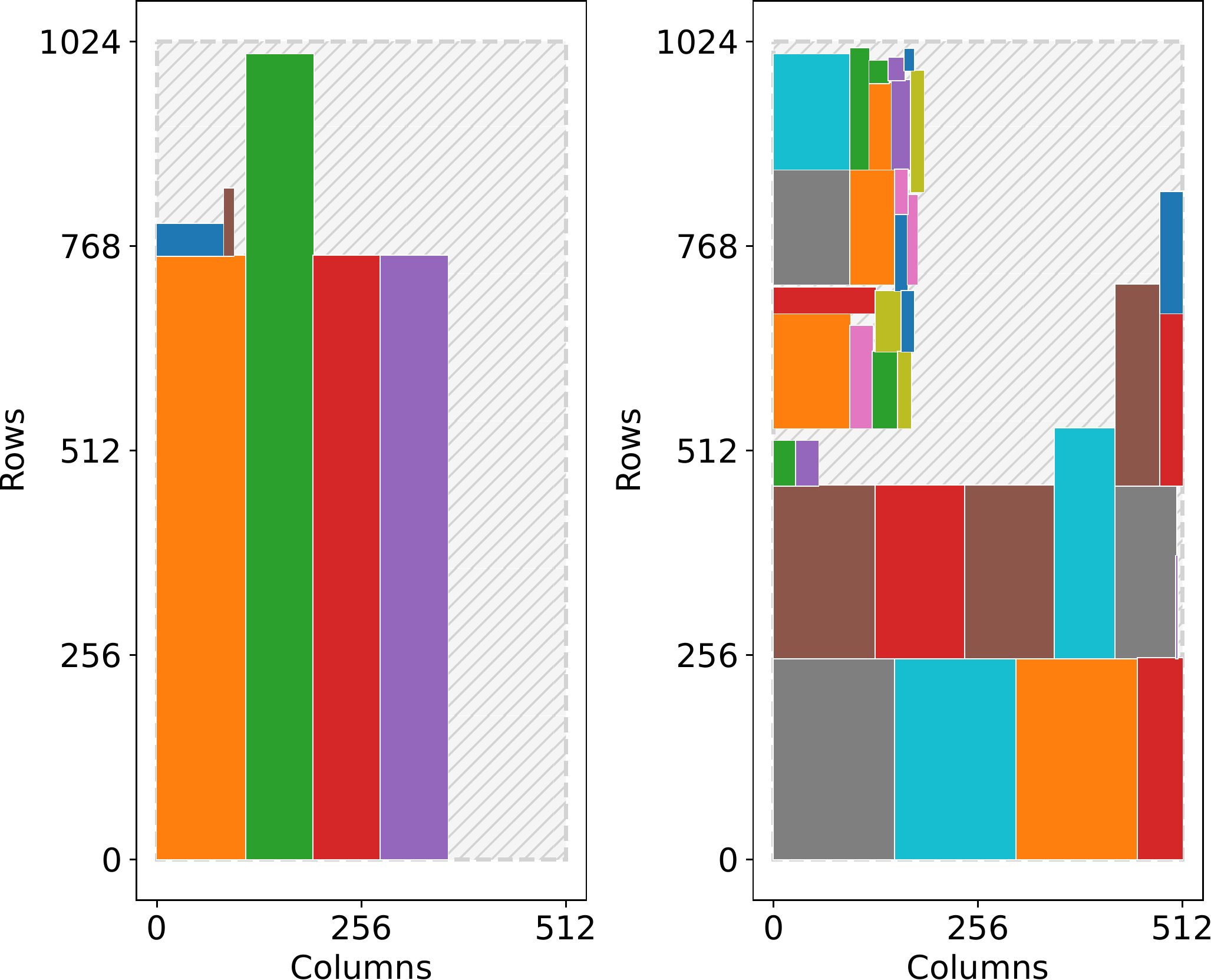}
    \vspace{-0.2cm}
    \caption{
    Mapping to CiM array: (Left) AnalogNet-KWS (57.3\% utilization), and (Right) AnalogNet-VWW (67.5\% utilization).
    %(right) mapping to the 1024$\times$512 crossbar with differential columns, 
    Each colored box represents a unique layer automatically mapped. 
    %achieving 57.3\% and 67.5\% utilization, respectively.
    }
    \label{fig:mapping}
\vspace{-0.4cm}
\end{figure}

\subsection{Layer-Serial Approach}

The conventional approach to mapping DNNs to CiM arrays is to map weights for each layer into a unique array, and then stream the activations through the arrays in a parallel/fully-pipelined style~\cite{Y2021dazziTC}.
However, while this achieves the highest possible throughput, for TinyML applications, the fully-pipelined approach incurs area, power and control complexity overheads.
%is not area and power efficient.
In particular, two of these overheads are significant.
%there are a few significant area overheads.  
First, fully-pipelined requires a large, high-bandwidth interconnect between the individual arrays.
%This interconnect introduces \RED{XXX} area, energy and latency overheads.
%For programmable CiM, 
The interconnect is challenging to design, because the bandwidth required depends on the model architecture~\cite{Y2021dazziTC}.
%Although these functions are fairly cheap to compute, if they are required on each array, they can inflate the area rapidly.
Second, layer-parallel requires each layer to be in a different CiM array.  This increases the number of arrays and correspondingly the number of ADCs and DACs.
%increases.
Each array also must incorporate the compute required in between convolutions, such as activation functions, scaling, pooling etc, further increasing area.

We propose an alternative approach of processing the network in a \textit{layer-serial} fashion:
process one layer at a time until it is complete, after which we start the next layer.
%There is no dependency on the latency of a given layer, as we will not start the next layer until the current one is completed.
Since only the weights associated with a single layer are used in any cycle, we do not need separate arrays for each layer, and we can store the whole model weights in the same CiM array.
To illustrate this, Figure~\ref{fig:mapping} gives mappings of the two AnalogNets models onto a single CiM array, with each layer represented by a colored box.
With a single large array, the periphery cost of the DACs and ADCs is more heavily amortized, leading to improved energy efficiency.
Furthermore, a programmable interconnect is no longer required, because the activations simply circulate from the output of the array to SRAM and from the SRAM to the input of the array.
%later back to the input of the array in the subsequent layer.

\subsection{Accelerator Architecture}
\label{sec:hw_arch}

Figure~\ref{fig:accelerator} gives the AON-CiM accelerator architecture, consisting of two main parts: 1) PCM CiM array, and 2) the activation processing and storage pipeline.
% A thorough description of the architecture design space exploration and design choices is described in Section~\ref{sec:appendix_hw} in the Appendix.

\paragraph{Analog CiM Array}
The array itself is based on~\citet{Y2021khaddamaljamehVLSI}, but with 1024 rows and 512 columns to fit either of the AnalogNets models with some headroom for programmability.
The tall aspect ratio is desirable, as ADCs consume more area than DACs, and besides, AnalogNets-KWS has tall layers, as shown in the mappings in Figure~\ref{fig:mapping}.
%Figure~\ref{fig:mapping} shows the mapping of both models onto the array.
Since we only use a subsection of the array for any given layer, we will rarely use all the DACs or ADCs at once.
Therefore, to save power, we clock gate unused DACs and ADCs.
%and multiplex the required data to the DAC inputs and from the ADC outputs.
% 3.39952192E-06 vs 3.19984192E-06
To save area, we use a 4-input analog multiplexer on the bit-lines for a 4$\times$ reduction in ADCs (6\% area benefit).
We support 8/6/4-bit activation precision, as the CiM cycle time drops significantly due to the PWM DAC, which has exponential latency with bitwidth.

\paragraph{Activation Processing and Storage}
Besides matrix-vector operations on the CiM array, we must also implement a number of mostly vector-wise operations on the pre-activation outputs from the array, including scaling, batch normalization, ReLU, pooling etc.
%These operators include scaling, batch normalization, ReLU, pooling etc.
The nominal 8-bit peak activation processing throughput required for this is 128 data words per 130ns array cycle, each of which requires two floating-point scalings and various integer functions depending on the layer type.
However, supporting 4-bit activations without stalling the array requires the same 128 data words throughput at a 10ns array cycle time.
%case is even more severe, as the array cycle time reduces exponentially to 
%10ns, while requiring the same throughput from the activation processing.
%per quantized activations, the \RED{512} INT8 words would involve \RED{1024} FP ops --Figure~\ref{fig:accelerator}.
Therefore, we add a digital datapath to meet the worst case 4-bit throughput,
%to reduce the area required for this activation processing, we exploit the relatively long cycle time of the array (130ns for 8-bit operation) 
%, to reduce the area by running the activation processing 
%by 
using an 800 MHz clock.
%for digital processing to minimize datapath width and save area.
%, which is $104\times$ faster than the array latency.
%This allows us to reduce the activation datapath width by a corresponding factor to \RED{$\lceil 1024/(130ns/1.25ns) \rceil =10$}. 
%Instead of implementing a \RED{1024}-wide datapath we take advantage of the faster clock frequency, reducing the number of required multiplier units to $\lceil 1024/(130ns/1.25ns) \rceil =10$ multiplier units, meeting the timing requirements and without increasing the latency or energy. 
%In this work we carried out a design space exploration of how the number of parallel FP multiplier units determine the final area, throughput and energy efficiency under real workloads. The study results, detailed in Figure \ref{fig:ppa_fp_exploration} in the Appendix, show how by 32 multipliers significantly improve the system throughput at lower activation precision, with a negligible area impact.
Finally, the CiM input is fed with activations from the previous layer, stored in a double-buffered 128KB SRAM.
A hardware IM2COL unit expands the activations before they are applied to the DACs in the CiM, using a small buffer and a programmable address generator.
The activation processing, SRAM write/read and IM2COL are all pipelined, such that the array is never stalled (Figure~\ref{fig:accelerator}), even in the challenging 4-bit case.
%Figure~\ref{fig:operation} describes the different modules involved during the inference process.
%\todo{cite hw im2col patent}
%The SRAM is 128KB in total, and arranged as two logical banks such that output activations from the current layer can be written, while input activations from the previous layer are read in the same cycle.
Details of the control plane are omitted for brevity.
%\todo{describe DMA, AGU and control flow briefly if there's space}

\section{Results}
\label{sec:results}

\subsection{Methodology}

\paragraph{Model setup and training}
%For model training, 
We follow the data preprocessing and training setup of~\citet{banbury2021micronets} to train baseline AnalogNet-KWS on the Google Speech Commands (V2) dataset and baseline AnalogNet-VWW on the Visual Wake Words dataset (100$\times$100 resolution). 
We also trained the same architectures using our HW-informed training methodology (Section \ref{sec:train_methodology}). 
In the first training stage, AnalogNet-KWS (AnalogNet-VWW) was trained for 100 (200) epochs, with cosine learning rate decay and weight clipping. 
The second training stage uses the same numbers of epochs and learning rate (LR) decay schedule, but the initial LR is reduced to $1/10$ of its value in the first stage. An LR for the quantizer ranges is also introduced, with an exponentially decay from $10^{-3}$ to $10^{-4}$.
%during training. 
A gradient clipping threshold of $0.01$ is placed on the gradient for $S$ to stabilize its gradient update. We also exploit the stochastic ``quantization noise"~\cite{fan2020training} method to accelerate model training convergence at low bitwidths. 
The ``quantization noise" probability of quantizers is set to $0.5$.
% \todo{If we need the space we can move some of this paragraph to the appendix...}

\paragraph{Accuracy Evaluation}
After training the models, the weights (clipped to [$W_{l,\mathrm{min}}$, $W_{l,\mathrm{max}}$] and with no noise added), and the ranges of quantizers modelling the ADCs and DACs are transferred to a calibrated PCM CiM simulator, 
%calibrated by real device measurement, 
and further tested on a real 90nm PCM chip. 

In the simulator, the clipped weights are rescaled to $[-1,1]$ by dividing $W_l$ by $\textnormal{max}(|W_l|)$ and split into two arrays of equal size representing the positive and negative parts in conductances, which we denote as the target conductances $G_T$. After rescaling, the programming noise is simulated. The programmed conductances, denoted by $G_P$, are modeled using $G_P = G_T + \mathcal{N}(0,\sigma_P)$ where $\sigma_P=\textnormal{max}(-1.1731 G_T^2 + 1.9650 G_T + 0.2635,0)$. After the conductances are programmed, they drift over time, which also must be accounted for in the simulation~\cite{krebs_drift}. Drift at time $t$, given the initial time of programming $t_c=25s$, can be modelled using $G_D=G_P(t/t_c)^{-\nu}$, where $\nu$ is the drift coefficient that follows a normal distribution. Finally, when a matrix-vector multiplication is performed there will be instantaneous fluctuations on the hardware conductances due to the intrinsic noise from the PCM devices. PCM exhibits $1/f$ noise and random telegraph noise characteristics, which alter the effective conductance values used for computation. The final conductances in simulation follow a normal distribution $G \sim \mathcal{N}(G_D,\sigma_{nG}(t))$, where $\sigma_{nG}(t)=G_D(t) Q \sqrt{\textnormal{log}((t+t_r) / t_r)}$ with $t_r=250\mathrm{ns}$ and $Q=\textnormal{min}(0.0088/G_T^{0.65},0.2)$.

\paragraph{Accuracy Experiments on Prototype PCM Hardware}
Furthermore, we test model accuracy on a PCM-based prototype hardware chip, fabricated in 90 nm CMOS technology node~\cite{Y2010closeIEDM}. PCM devices are integrated into the chip via a key-hole process and doped Ge$_2$Sb$_2$Te$_5$ is used as the phase-change material. The PCM array is organized as a crossbar with 512 source lines and 2048 bit lines, and an NMOS transistor serves as the PCM access device. The source lines and bit lines can only be serially addressed. A close-loop programming algorithm, as described in~\cite{Y2020joshiNatComm} is used to program the devices. A fixed voltage of 300 mV amplitude is applied to the source lines.
%for the read operation. 
The sensed current is integrated on a capacitor and the voltage is then converted to digital by 8-bit on-chip ADCs.

\paragraph{Hardware power, performance and area}
The AON-CiM accelerator was modeled in a 14nm process technology.
The digital part was evaluated using 14nm cell libraries and SRAM compilers, while the 14nm PCM CiM array was based on measurements reported by \citet{Y2021khaddamaljamehVLSI}.
A bespoke layer-serial layer tiler was developed, along with a cycle accurate simulator.
Together, these allow us to automatically map a given model to the AON-CiM accelerator and evaluate latency and energy.

\subsection{AnalogNets Accuracy on PCM CiM Simulator}

\begin{figure*}[t!]
\begin{center}
\includegraphics[width=0.33\textwidth]{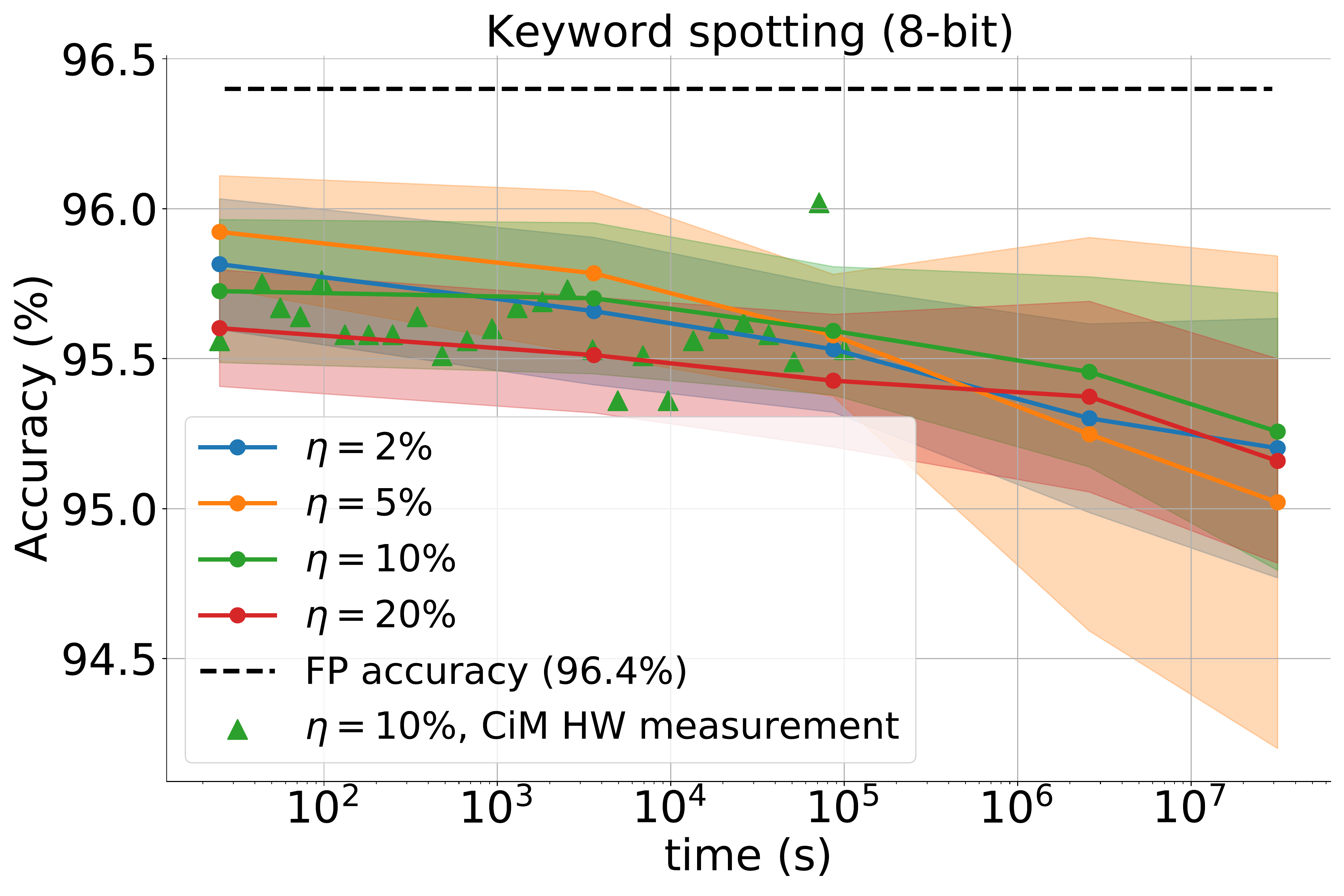}
\includegraphics[width=0.33\textwidth]{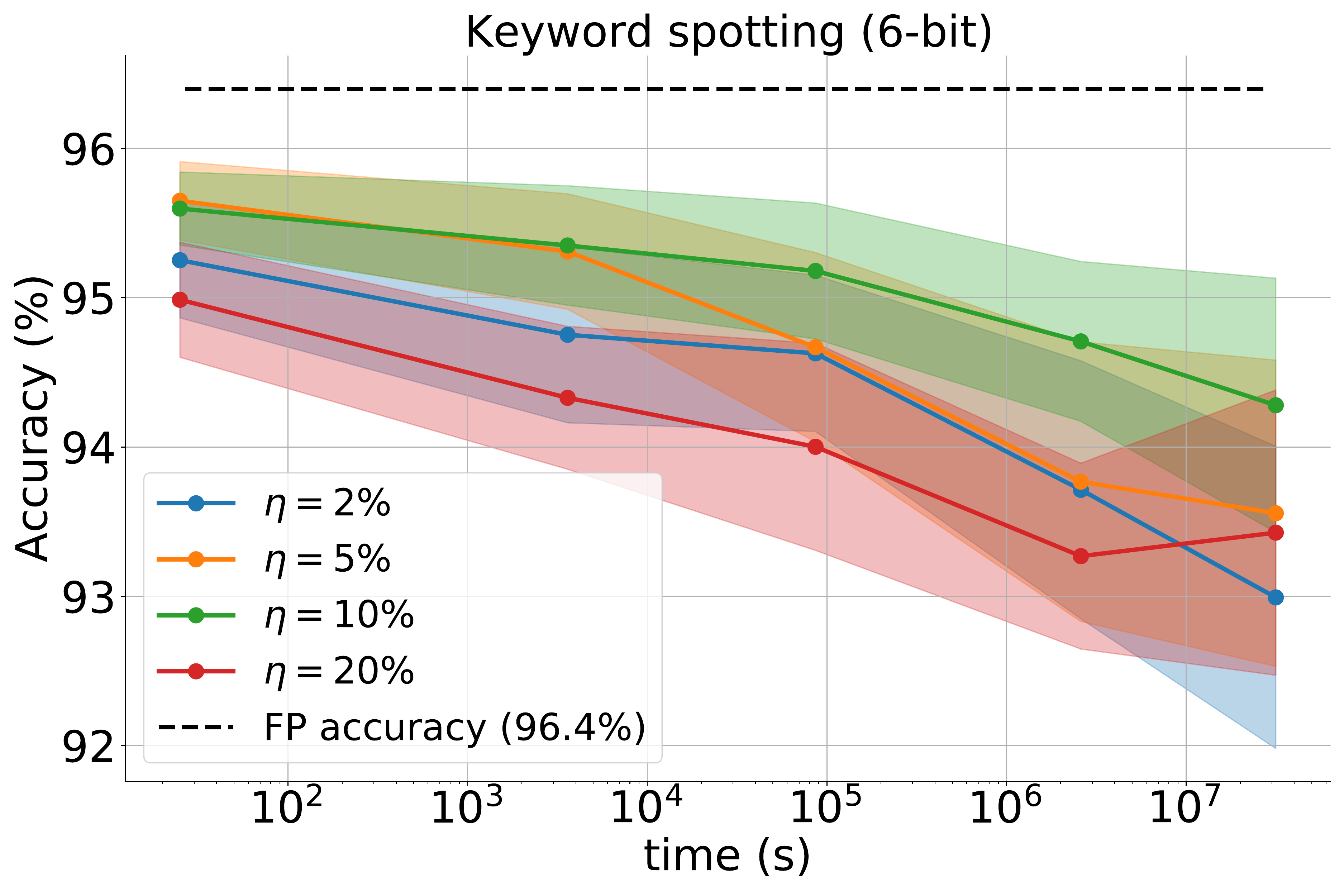}
\includegraphics[width=0.33\textwidth]{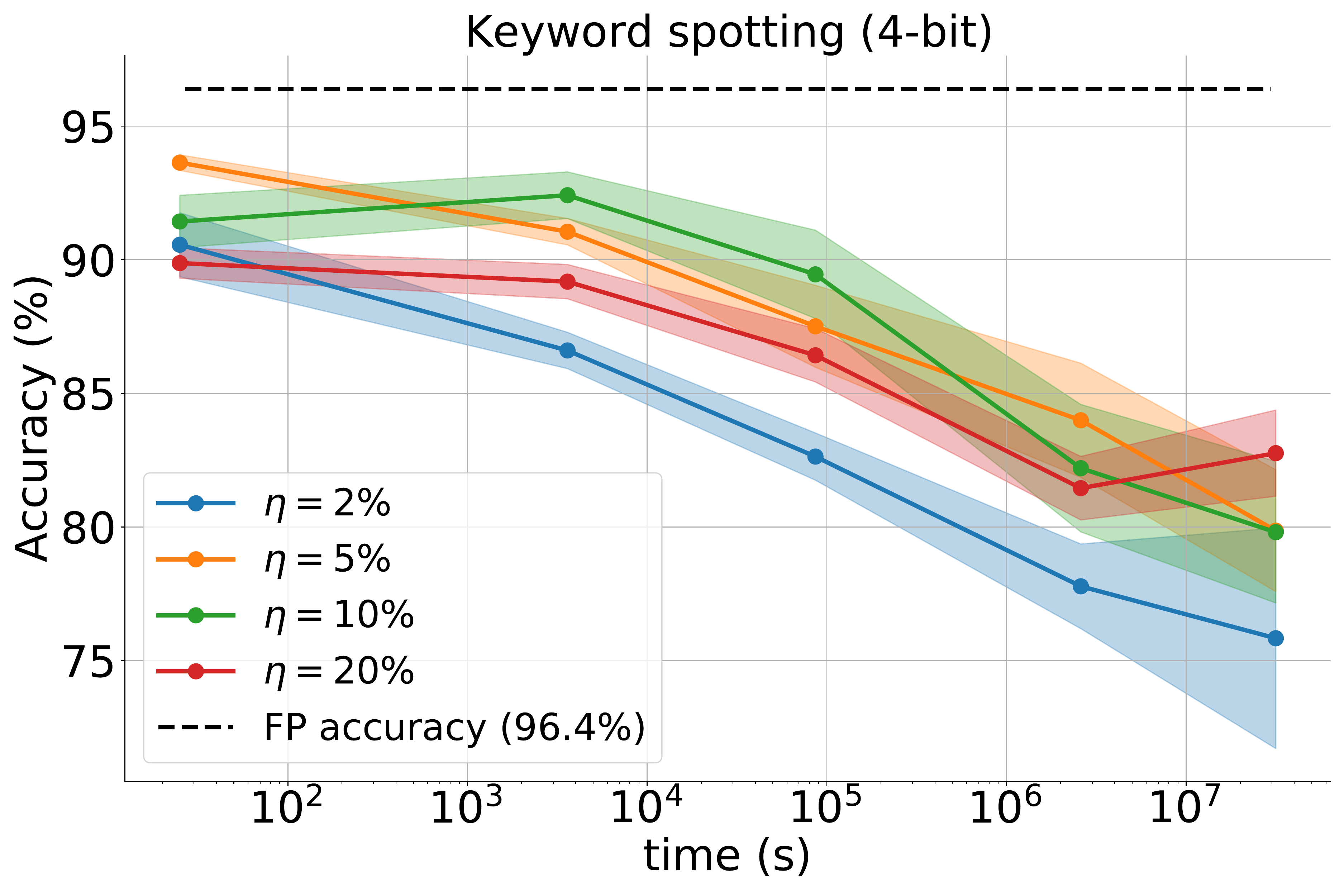}
\includegraphics[width=0.33\textwidth]{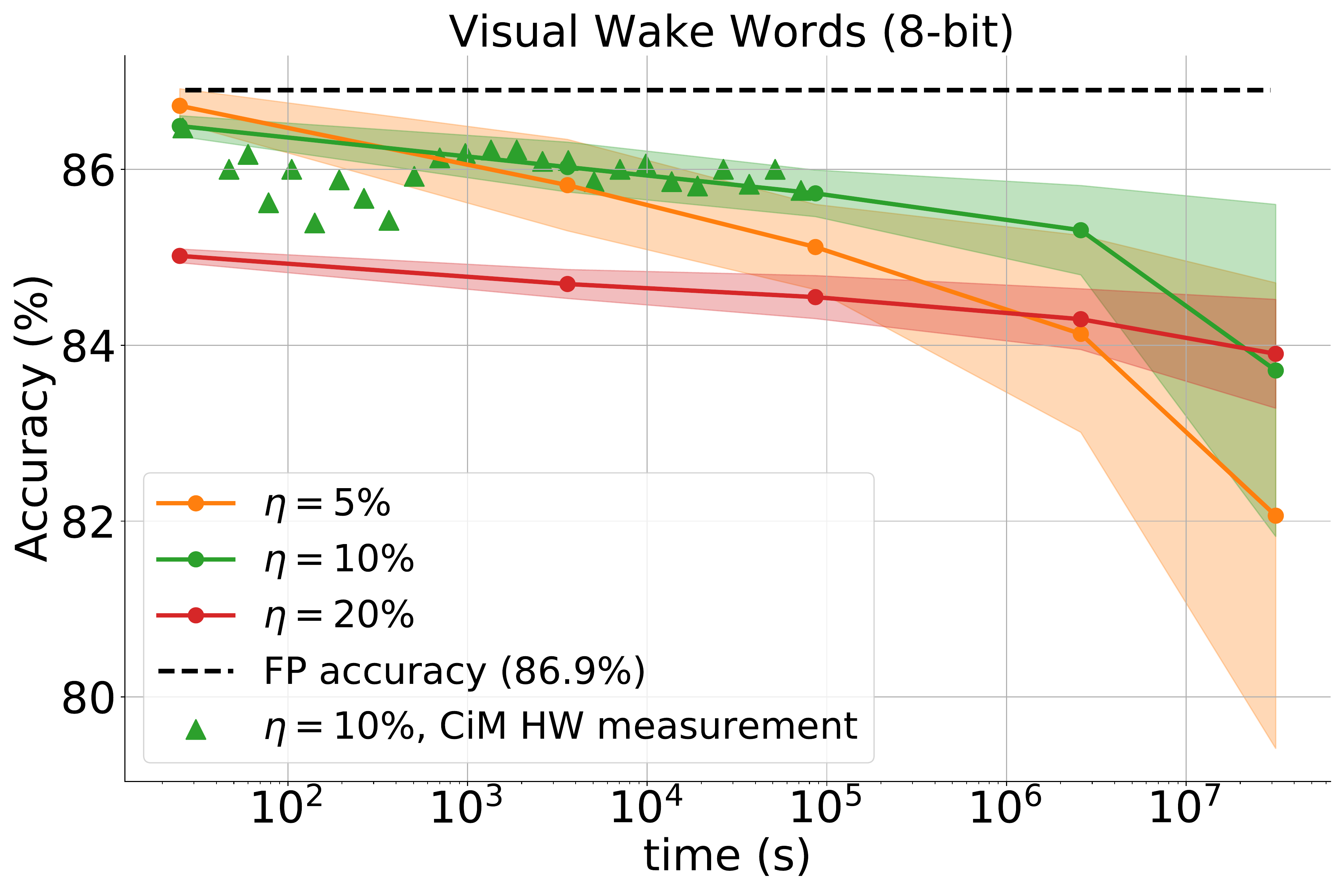}
\includegraphics[width=0.33\textwidth]{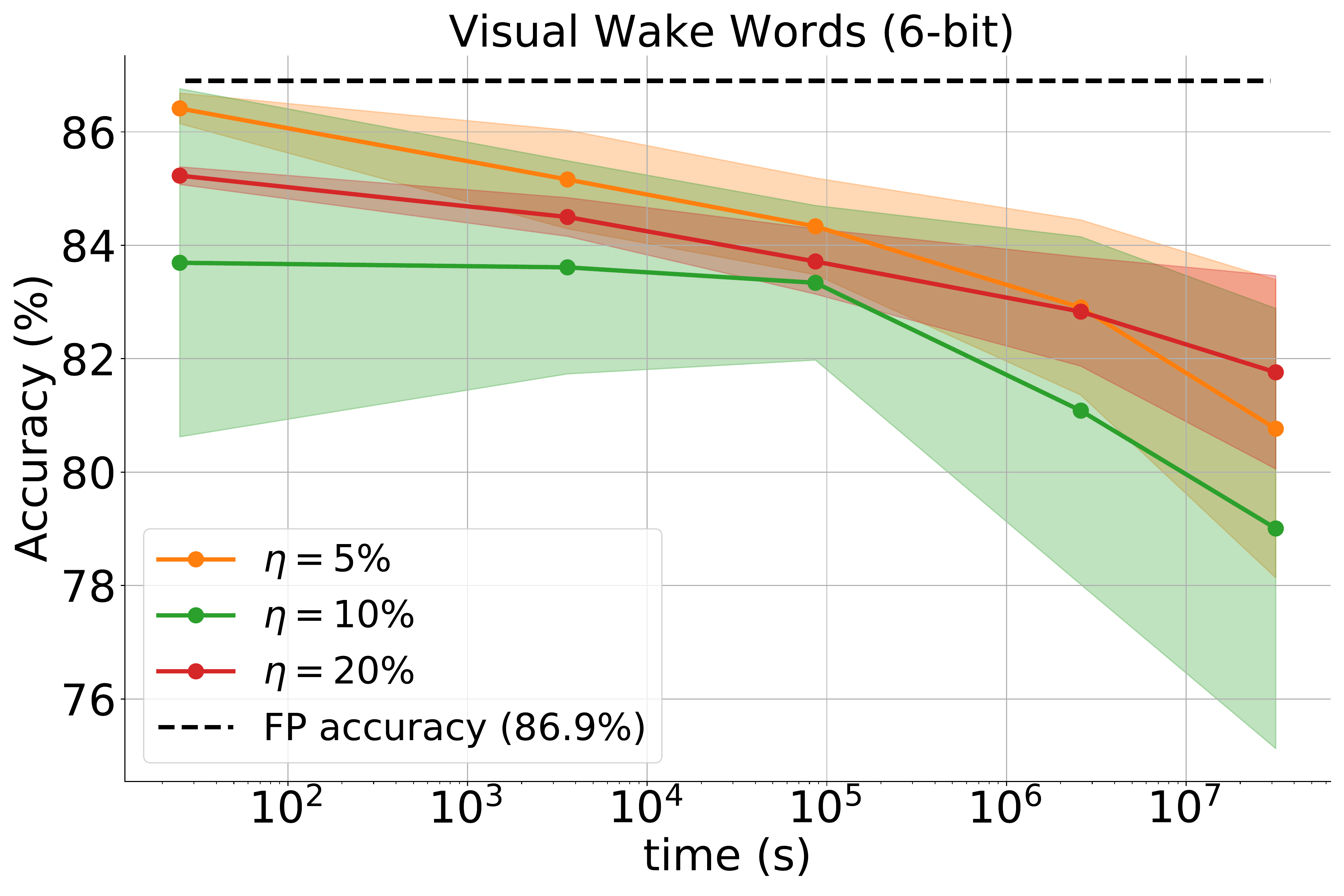}
\includegraphics[width=0.33\textwidth]{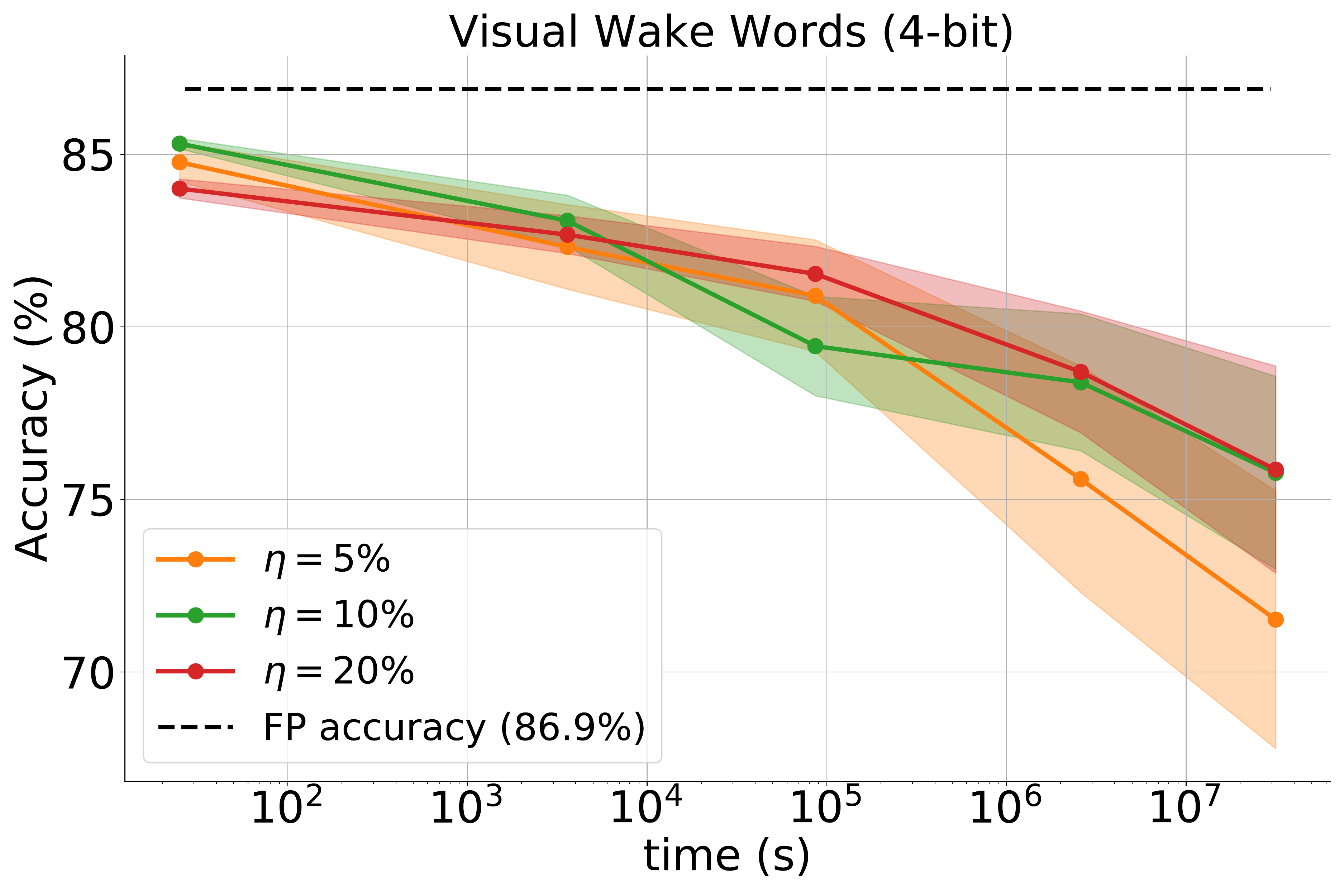}
\end{center}
\vspace{-0.5cm}
\caption{
Accuracy of AnalogNet-KWS and AnalogNet-VWW in simulation and measured on real PCM hardware. The uncertainty bands represent the estimate of one standard deviation. $\eta$ describes the level of noise injected during model training. The triangles are data points from PCM hardware measurement during 20-hour runs (up to 24 hours for KWS).}\label{fig:accuracy}
\end{figure*}

Models trained at varying noise levels were deployed on the simulator, configured with a 1024$\times$512 CiM array, such that no layers are split.
%in these models. 
%to charaterize their accuracy.
%after some deployment time. 
Conductance drift in the PCM weights causes the model accuracy to degrade over time, even with the global drift compensation (GDC)~\cite{Y2020joshiNatComm}. 
%It is therefore important to look at a time series of accuracy data. 
The simulator calculates accuracy after 25 seconds, 1 hour, 1 day, 1 month and 1 year of deployment. 
The means and standard deviations are computed from 25 runs.

\subsubsection{Impact of Training Noise}
We first explore sensitivity to training noise injection level $\eta$, which is a training hyperparameter.
%of the problem. 
The optimal value of $\eta$ may depend on the exact HW characteristics and the model itself. We show results for $\eta \in \left\{2\%, 5\%, 10\%, 20\%\right\}$ on KWS and for $\eta \in \left\{5\%, 10\%, 20\%\right\}$ on VWW. 
Figure \ref{fig:accuracy} summarizes the model accuracy performance over time for AnalogNet-KWS and AnalogNet-VWW at different activation precision. 
%For example, a 6-bit model is trained and deployed with a configuration of 7-bit ADC and 6-bit DAC. 
For AnalogNet-KWS, the best training noise level $\eta$ is around $10\%$, while for AnalogNet-VWW, it is closer to $20\%$, especially for lower activation bitwidths.

\subsubsection{Quantization}
%The simulation results are plotted on logarithmic scale. 
On a logarithmic timescale, accuracy drops more sharply when the activation precision is reduced. 
On KWS, accuracy is maintained within $2\%$ of the digital floating point baseline for a year at 8-bit. 
While with a 6-bit ADC, that time is reduced to about a month. 
At 4-bit, the model can maintain $90\%$ accuracy for only about a day. 
On VWW, an accuracy drop of less than $3\%$ can be maintained for a year with 8-bit ADCs. 
At 6-bit, the $3\%$ accuracy drop can be maintained for about one month. 
At 4-bit, the model stays above $80\%$ for about a day. 
As we will see next,
%n the following sections, 
activation bitwidth has a big impact on latency and energy efficiency on the AON-CiM accelerator. 
Therefore, one could use activation precision to trade-off accuracy with hardware performance.
%depending on its accuracy / performance trade-off.

\subsubsection{Ablation}
Table~\ref{tab:ablation} further shows an ablation study of the simulator results, demonstrating the advantage of our end-to-end training method.
%proposed in this paper. 
We compare with off-the-shelf models without special re-training, which drop to almost random level accuracy when tested on the simulator, and vanilla noise injection training~\cite{Y2020joshiNatComm}.
Vanilla noise injection can yield results close to our method at 8-bit activation precision, but becomes worse at lower precision.
We also compare with the VWW model that has the small bottleneck layers added back (last row). 
The two models are trained in the same way with noise injection and ADC/DAC modeling.
The VWW model with bottlenck layers, despite having more parameters than AnalogNet-VWW, has worse accuracy. 
Our end-to-end training methodology with ADC/DAC constraints also gives optimized ready-to-use 
%quantities for the 
DAC scaling parameters and ADC gains, which would otherwise need to be computed by sub-optimal empirical rules (see Appendix).
%for more details). 

\begin{table}[t]
\caption{Accuracy ($\%$) after 24 hours PCM drift (simulation) for different training methods and design choices. Our approach shows consistently better results especially at low activation bitwidth.} 
\label{tab:ablation}
%\resizebox{1.\linewidth}{!}{%
\footnotesize
\begin{tabular}{|l||c|c|c|}
\hline
\multicolumn{4}{|c|}{\textbf{KWS}}  \\ \hline
Activation bitwidth
 & 8bit & 6bit & 4bit \\ \hline         
 Baseline (no re-training) & \begin{tabular}[c]{@{}c@{}}9.4 \\+/- 0.6\end{tabular} & \begin{tabular}[c]{@{}c@{}}9.4\\ +/- 0.6\end{tabular} & \begin{tabular}[c]{@{}c@{}}8.6\\ +/- 0.1\end{tabular}  \\ \hline
Noise injection ($\eta=10\%$) & \begin{tabular}[c]{@{}c@{}}95.4\\ +/- 0.3\end{tabular}                                   & \begin{tabular}[c]{@{}c@{}}85.0\\ +/- 1.1\end{tabular}                                   & \begin{tabular}[c]{@{}c@{}}15.1\\ +/- 1.4\end{tabular}    \\ \hline
\begin{tabular}[c]{@{}c@{}} Noise injection ($\eta=10\%$) \\
and ADC/DAC constraints \end{tabular} & \textbf{\begin{tabular}[c]{@{}c@{}}95.6\\ +/- 0.2\end{tabular}}                         & \textbf{\begin{tabular}[c]{@{}c@{}}95.2\\ +/- 0.5\end{tabular}}                          & \textbf{\begin{tabular}[c]{@{}c@{}}89.5\\ +/- 1.7\end{tabular}}                                      \\ \hline
\multicolumn{4}{|c|}{\textbf{VWW}} \\ \hline
Baseline (no re-training) & \begin{tabular}[c]{@{}c@{}}74.6\\ +/- 3.5\end{tabular} & \begin{tabular}[c]{@{}c@{}}52.6\\ +/- 0.2\end{tabular} & \begin{tabular}[c]{@{}c@{}}52.8\\ +/- 0.2\end{tabular}   \\ \hline
Noise injection ($\eta=10\%$) & \begin{tabular}[c]{@{}c@{}}85.7\\ +/- 0.4\end{tabular} & \textbf{\begin{tabular}[c]{@{}c@{}}83.7\\ +/- 0.5\end{tabular}} & \begin{tabular}[c]{@{}c@{}}52.4 \\ +/- 0.2\end{tabular}    \\ \hline
\begin{tabular}[c]{@{}c@{}} Noise injection ($\eta=10\%$) \\
and ADC/DAC constraints \end{tabular} & \textbf{\begin{tabular}[c]{@{}c@{}}85.7\\ +/- 0.3\end{tabular}} & \begin{tabular}[c]{@{}c@{}}83.3\\ +/- 1.4\end{tabular} & \textbf{\begin{tabular}[c]{@{}c@{}}79.4\\ +/- 0.14\end{tabular}} \\ \hline
 Bottleneck layers included &
\begin{tabular}[c]{@{}c@{}}83.0\\ +/- 1.2\end{tabular} &
\begin{tabular}[c]{@{}c@{}}71.7\\ +/- 6.2\end{tabular} &
\begin{tabular}[c]{@{}c@{}}54.1\\ +/- 1.0\end{tabular} \\ \hline
\end{tabular}
%}
\centering
\end{table}

\subsection{AnalogNets Validation on PCM CiM Test Chip}

To simulate the model using CiM hardware, the weights are initially scaled to target conductances, which are then serially programmed into the hardware. After the conductances have been fully programmed, we wait for 20 hours while serially reading out the conductances at exponentially spaced points in time. The obtained conductances are then scaled down to the original weight magnitudes and used in the simulator to obtain the test accuracy. The results obtained using the PCM hardware are shown in Figure \ref{fig:accuracy}. The experiments are nicely aligned with the model simulations.

There are minor discrepancies between the simulator and the PCM hardware experiment. One reason for this is that our simulator does not take into account the convergence of the iterative programming algorithm. Although the overall convergence is above 99\% for both models, a slightly lower convergence ($\sim$98.5\% ) is recorded for larger absolute wight values. Moreover, only a single experiment is evaluated while the simulations are repeated for 25 runs. 

\subsection{AON-CiM Accelerator Evaluation}

\begin{figure}[t!]
    \centering
	\includegraphics[width=\columnwidth]{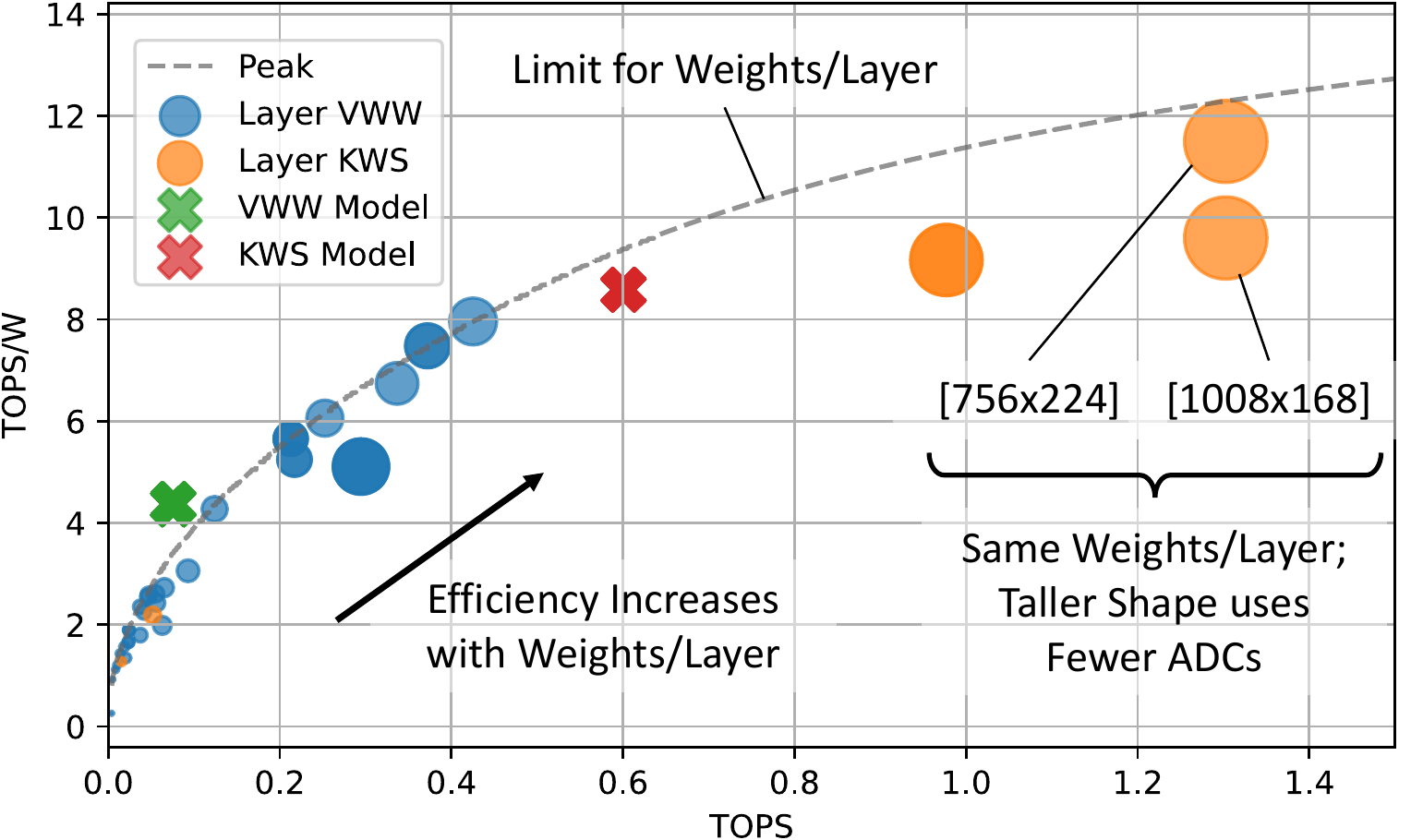}
    \vspace{-0.8cm}
    \caption{
    Layer/model-wise AON-CiM accelerator efficiency and throughput for the two AnalogNet models with 8-bit activations. Marker size indicates the number of weights in the layer.
    %decomposition for both analyzed workloads after its deployment on a single 1024$\times$512 differential crossbar assisted by $1024$ DACs and $128$ ADCs and $32$ FPu units.
    }
    \label{fig:tops-topsw-for-layers}
\end{figure}

Table~\ref{table:hw-summary} summarizes the 14nm AON-CiM accelerator, including throughput and energy efficiency on the AnalogNets models.
At 100\% utilization, 
%We first characterize the maximum performance achievable, corresponding to a single Fully-Connected or 2-dimensional convolutional layer whose GEMM operation utilizes $100\%$ of the crossbar, maximizing the used resources.
%As summarized in the table, the throughput strongly depends on the DAC/ADC precision, determining this data-converters periphery the mean-power and the CiM module latency. 
the single large CiM array achieves peak performance of
%sing the presented single tile accelerator, 
$2/7.71/26.21$ TOPS and $13.55/45.55/112.44$ TOPS/W at 8/6/4-bit activation precision, respectively.
The major gains at smaller bitwidths are due to the latency of the PWM DAC, which scales exponentially with bitwidth.
%In the same manner, 
%The peak energy efficiency varies from $30-390$ TOPS/W based on periphery precision.
%However, and as discussed in Section~\ref{sec:models}, the number of parallel operations is limited by the geometry of the accelerated GEMM operations. 
%A clear trend between the number of parallel \emph{ops} (defined as either a multiplication or an add) and the crossbar size is established.
%\RED{
%Do we need this?
%The throughput and efficiency numbers achieved on smaller arrays is limited by the maximum height of the array where the layer can be deployed --Figures~\ref{fig:ppa_256_0},~\ref{fig:ppa_256_1} in Appendix. 
%Such limitations are clearly seen in the mappings depicted in Figures~\ref{fig:mapping_kws_256} and ~\ref{fig:mapping_vww_256}.
%}
However, real models do not use the whole array at once, so the achievable efficiency on a real model depends on
%is a function of both the hardware (including pre-activation processing and SRAM read/write) and 
the DNN architecture itself.
%(i.e. the layer sizes and shapes).
%, as the layer sizes and shapes determine.
%These ideal peak throughput/efficiency are hindered the moment the model layers do not meet the optimal proportions.
%For both the workloads analyzed in can be appreciated how the workload throughput sets at 

Figure~\ref{fig:tops-topsw-for-layers} shows TOPS and TOPS/W for both individual layers and whole model performance of AnalogNets executing on the AON-CiM accelerator.
We highlight two interesting trends.
First, larger layers (shown by marker size) achieve higher TOPS and TOPS/W, because the DAC/ADC cost is amortized over more MACs.
Second, even for layers with the same size, those with a tall aspect ratio have higher TOPS/W.
This is because ADCs consume more energy than DACs, and taller layers require less ADCs for the same number of MACs.
The limit on this aspect ratio benefit is shown 
%maximum throughput/efficiency limit for a given layer size is shown i
as a dotted line for comparison, which some smaller layers meet.
%It can be seen how taller layers, represented with larger markers correlate with higher throughput and efficiencies.
AnalogNet-KWS has a small number of tall layers, and hence achieves higher TOPS and TOPS/W than the smaller AnalogNet-VWW layers.
Overall, the two models achieve throughput of $0.6/2.29/7.8$ and  $0.076/0.29/0.98$ TOPS for KWS and VWW respectively, at $8/6/4$ bits activation precision.
Their energy efficiency is $8.58/26.76/57.39$ and $4.37/12.82/25.69$ TOPS/W for KWS and VWW respectively, at $8/6/4$ bits activation precision.

\subsection{Discussion}
Finally, we briefly discuss our results in comparison with related work on TinyML for both analog CiM and MCUs.

\paragraph{TinyML on Analog CiM} 
Compared to previous work, AnalogNets demonstrates significantly higher accuracy.  For example, on the KWS task, \citet{dbouk-kws-jssc2021} demonstrated a 65nm SRAM CiM test chip with 90.38\% accuracy (91K parameters) on just 7 keywords.
\citet{guo-rnn-vlsi2019} also report a 65nm SRAM CiM, which achieves 90.2\% accuracy on 10 keywords.
In contrast, AnalogNet-KWS achieves 95.6\% (after 24h drift) on the full 12 keywords. This accuracy gap highlights the lack of research on model architectures and training methodologies for analog CiM.

\paragraph{TinyML on MCUs}
MCUs are currently the commodity platform for TinyML.
On the same KWS and VWW tasks, \citet{banbury2021micronets} report 48.6 mJ/inf. at 95.3\%, and 196.2 mJ/inf. at 86.4\%, respectively.
% NOTE: These are MicroNet-KWS-S and MicroNet-VWW-3
While AnalogNets on AON-CiM accelerator show 8.2 uJ/inf. at 95.6\% and 15.6 uJ/inf. at 85.7\%, respectively.
This is a very significant four orders of magnitude improvement in energy efficiency at comparable task accuracy.
%However, while deploying models on MCUs is relatively straightforward, even with INT8 quantization, significant additional effort was required for the same task on analog CiM (Section~\ref{sec:models}).
%So, while the bespoke models and elaborate training flows we used for AnalogNets are unlikely to be adopted for mainstream applications, we believe that for severely energy constrained applications such as always-on TinyML, the effort may be justified.
Therefore, although analog TinyML requires more complex model design and training, which is unlikely to be adopted for mainstream application, our work suggests that for severely energy constrained applications such as always-on TinyML, the effort is justified.

\paragraph{Future Work}
%This work focuses on noise-robust, compact DNNs for analog CiM.  
%However, 
Figure~\ref{fig:tops-topsw-for-layers} shows that the size and shape of the layers has a huge impact on throughput and efficiency.
Therefore, it may be possible to directly optimize the layer shapes and sizes, without increasing the overall model size, to attempt to achieve higher energy efficiency on the same AON-CiM hardware at similar accuracy.
%to maximize throughput and energy efficiency.

\begin{table}[t]
\centering
\caption{
AON-CiM accelerator summary.
%on AnalogNets.
}
% \resizebox{0.33\textwidth}{!}{
\small
%\scriptsize
\footnotesize
\begin{tabular}{|>{\bfseries} l || c |}
\hline
Technology & 14nm CMOS
% , 5.34 $\mu$m$^2$ PCM Cell
\\ \hline
%Bitcell & Differential PCM (5.34 $\mu$m$^2$ cell) \\  \hline
Array Size & 1024 Rows $\times$ 512 Columns (Mux4) \\  \hline
%FP Activations & $32\times2.7$mW \\  \hline
%DAC & 1024 PWM DACs\\  \hline
%ADC & 256 ADCs \RED{4-input mux} \\ \hline
%FP FMA Units & $64$ \\ \hline
SRAM & 128 KB in Two Banks\\ \hline
Digital Ops & FP Scale, ReLU, Pooling, Add, IM2COL \\ \hline
%\multirow{2}{*}{Clock Period}
    %& $T_{CiM}$ (8/6/4b): 130 ns / 34 ns / 10 ns \\
Clock    & $T_{CiM}$: 130ns (8b), 34ns (6b), 10ns (4b) \\
    %& PWM : 1 GHz \\
Period    & $T_{Digital}$: 1.25ns \\ 
\hline
    %& SRAM: 200 MHz \\ \hline
Chip Area
    & 3.2mm$^2$ Total \\
    & 3.07mm$^2$ CiM, 0.15mm$^2$ Digital \& SRAM \\
    %& 3.07mm$^2$ (CiM), 0.153 mm$^2$ Digital & SRAM, 3.2 mm$^2$ (Total)\\
    %& Digital Processing 32$\times$1480$\mu$m$^2$ \\
    %& SRAM 0.106 mm$^2$ \\ 
%\multirow{4}{*}{Chip Area}
%    & PCM Array $2.8 mm^2$\\
%    & Periphery $0.27 mm^2$\\
%    & Digital Processing 32$\times$1480$\mu$m$^2$ \\
%    & SRAM 0.106 mm$^2$ \\ 
%\hline
%Total Area & $3.2mm^2$ \\
\hline
%\multirow{2}{*}{Throughput}
  Throughput & \textbf{KWS}: \quad 0.6 TOPS, \quad\quad 7,762 inf/sec \\
  (8b Act.)  & \textbf{VWW}: \quad 0.075 TOPS, \quad 1,063 inf/sec \\ \hline
%\multirow{2}{*}{Energy Eff.} 
Energy Eff.  & \textbf{KWS}: \quad 8.58 TOPS/W, \quad 8.22 $\mu$J/inf \\
(8b Act.)    & \textbf{VWW}: \quad 4.37 TOPS/W, \quad 15.6 $\mu$J/inf \\
\hline
\end{tabular}
\label{table:hw-summary}
\end{table}

\section{Conclusion}
\label{sec:conclusion}

Analog CiM hardware promises compelling improvements in energy efficiency, and is especially appealing for always-on TinyML tasks.
However, analog CiM introduces numerous new practical challenges, such as various circuit non-idealities. 
It is therefore essential to address these challenges in both the design of the DNN architecture and the training of the model.
In this paper, we described an ML-HW co-design approach to design AnalogNets: TinyML models optimized and trained specifically for analog CiM hardware, including aggressive quantization.
AnalogNets demonstrate accuracy close to the digital floating point baselines, when tested on both a calibrated simulator and a real PCM testchip.
We also described the AON-CiM accelerator, which introduces a novel layer-serial approach to minimize area for IoT applications.
The combined effort gives close-to-digital model accuracy on a significantly more energy efficient hardware platform.
%for these applications. 
%This work is a promising first step towards enabling deployment of always-on TinyML applications on analog CiM accelerators.

\if0
% \section*{Acknowledgements}

% \textbf{Do not} include acknowledgements in the initial version of
% the paper submitted for blind review.

% If a paper is accepted, the final camera-ready version can (and
% probably should) include acknowledgements. In this case, please
% place such acknowledgements in an unnumbered section at the
% end of the paper. Typically, this will include thanks to reviewers
% who gave useful comments, to colleagues who contributed to the ideas,
% and to funding agencies and corporate sponsors that provided financial
\fi

\section*{Acknowledgement}
This work was supported in part by the European Union's Horizon 2020 Research, Innovation Program through the project MNEMOSENE under Grant 780215.


\bibliography{papers}
\bibliographystyle{mlsys2021}

%%%%%%%%%%%%%%%%%%%%%%%%%%%%%%%%%%%%%%%%%%%%%%%%%%%%%%%%%%%%%%%%%%%%%%%%%%%%%%%
%%%%%%%%%%%%%%%%%%%%%%%%%%%%%%%%%%%%%%%%%%%%%%%%%%%%%%%%%%%%%%%%%%%%%%%%%%%%%%%
% SUPPLEMENTAL CONTENT AS APPENDIX AFTER REFERENCES
%%%%%%%%%%%%%%%%%%%%%%%%%%%%%%%%%%%%%%%%%%%%%%%%%%%%%%%%%%%%%%%%%%%%%%%%%%%%%%%
%%%%%%%%%%%%%%%%%%%%%%%%%%%%%%%%%%%%%%%%%%%%%%%%%%%%%%%%%%%%%%%%%%%%%%%%%%%%%%%

% added to clean figures floating
% \clearpage
\appendix
\clearpage
\section{Accuracy of MicroNet-KWS-S model on the PCM CiM simulator}
\begin{figure}[htb]
\begin{center}
\includegraphics[width=0.45\textwidth]{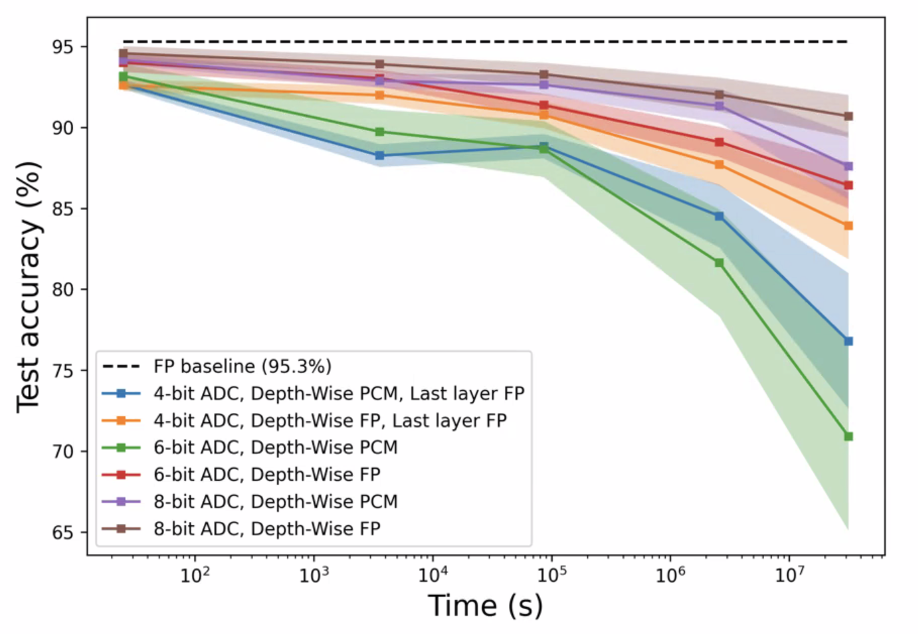}
\end{center}
\caption{Accuracy of MicroNet-KWS-S deployed on the simulator under different configurations and activation bitwidth. FP means floating point operations processed by a digital processor.}
\label{fig:micronet_acc}
\end{figure}

MicroNet-KWS-S model with depthwise separable convolutions is simulated on the PCM CiM simulator, results can be found in Figure~\ref{fig:micronet_acc}. 
With all the layers implemented in analog CiM, the accuracy (purple) dips to around $87.5\%$ after a year and implementing the depthwise convolutional layers in a digital processor can bring this accuracy above $90\%$ (brown), but it is still quite a bit worse than AnalogNet-KWS. 
Using lower activaiton bitwidth such as 6-bit and 4-bit shows even more severe detrimental effect of the depthwise layers on the model accuracy.

\section{Architecture of AnalogNet-KWS and AnalogNet-VWW}
\begin{figure*}[thb]
\begin{center}
\includegraphics[width=1.\textwidth]{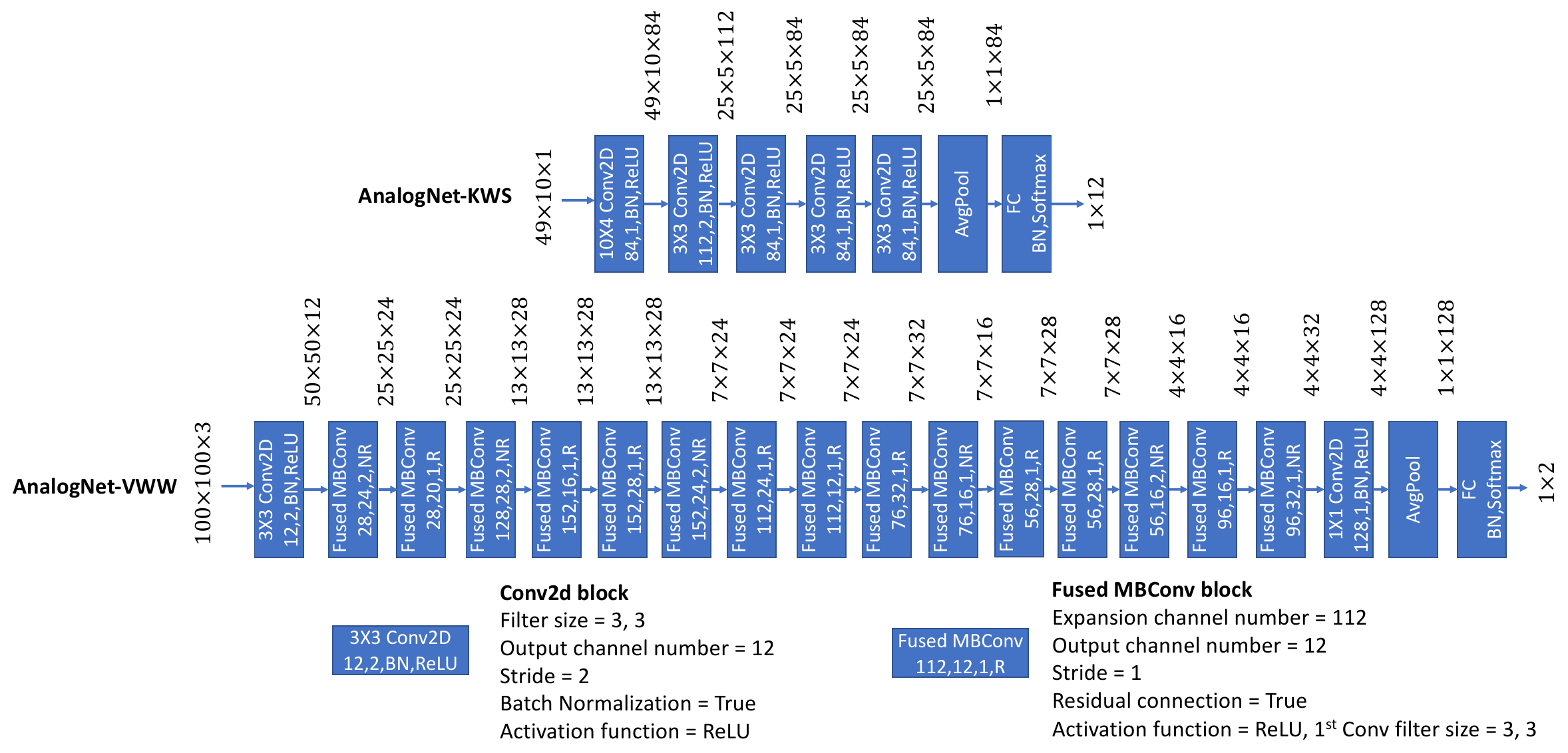}
\end{center}
\caption{Model architecture graph for AnalogNet-KWS (Top) and AnalogNet-VWW (Bottom) with a description of its building blocks.}
\label{fig:analognet_archi}
\end{figure*}

A detailed description of AnalogNet-KWS and AnalogNet-VWW model architecture can be found in Figure~\ref{fig:analognet_archi}. 

\section{Setting the DAC scaling factor and the ADC gain}
In the case when no trained DAC and ADC ranges are provided, the scaling factors $\mathrm{Scale}_{\mathrm{inp}}$ and $\mathrm{Scale}_{\mathrm{out}}$ (see Figure \ref{fig:HW2ML}) are calculated using heuristics. Because the CiM hardware that we use enables setting DAC ranges per layer, we calculated $\mathrm{Scale}_{inp}$ of the $l$-th layer as follows: $\mathrm{Scale}_{\mathrm{inp}}^l = (2^{n_{\mathrm{DAC}}-1}-1) / \mathrm{in}^l$, where $\mathrm{in}^l$ is the 99.995th percentile of the input activations for the $l$-th layer. The scale applied to the output going into the ADC can be calculated using

\begin{multline}
    \mathrm{Scale}_{\mathrm{out}}^l = ((2^{n_{\mathrm{ADC}}-1}-1) / n_{\textnormal{std-out}}) / ((2^{n_{\mathrm{DAC}}-1}-1)) \times \\
    G_{\mathrm{max}} \times \sqrt{\textnormal{size-crossbar}}) \times n_{\textnormal{std-in}} \times n_{\textnormal{w-std}},
\end{multline}
where $n_{\textnormal{std-out}}=n_{\textnormal{std-in}}=4.0$, $G_{\mathrm{max}}=25 \mu S$ and $\textnormal{size-crossbar}=1024$. When trained ADC ranges are provided, $\mathrm{Scale}_{\mathrm{out}}^l$ is calculated using $(2^{n_{\mathrm{ADC}}-1}-1) / \textnormal{trained}_{\mathrm{ADC}}$, with
\begin{multline}
\textnormal{trained}_{\mathrm{ADC}} = \frac{1}{L} \sum_{l=1}^{L}\textnormal{trained}_{\mathrm{ADC}}^l \times G_{\mathrm{max}} / \textnormal{max}(|W^l|) \\ \times (2^{n_{\mathrm{ADC}}-1}-1) / \textnormal{trained}_{\mathrm{DAC}}^l,
\end{multline}
where $W^l$ are the $l$-th layer weights, $\textnormal{trained}_{\mathrm{ADC}}^l$ is the trained ADC range of the $l$-th layer and $\textnormal{trained}_{\mathrm{DAC}}^l$ is the trained DAC range of the $l$-th layer. When trained DAC ranges are provided and because the input scale can vary per layer, we simply compute the input scale using $\mathrm{Scale}_{\mathrm{inp}}^l = (2^{n_{\mathrm{DAC}}-1}-1) / \textnormal{trained}_{\mathrm{DAC}}^l$.

\section{Depthwise Convolutional Layers: Deployment of MicroNet-KWS-S on a $1024 \times 512$ differential crossbar array}
\label{sec:dwcl_deployment}
\begin{figure*}[bp!]
    \centering
    \begin{minipage}{0.8\columnwidth}
    \centering
    \subfloat[MicroNet-KWS-S deployment on a $1024 \times 512$ NVM crossbar tile.]{\label{fig:depth_wise_kws:a}\includegraphics[width=0.98\columnwidth]{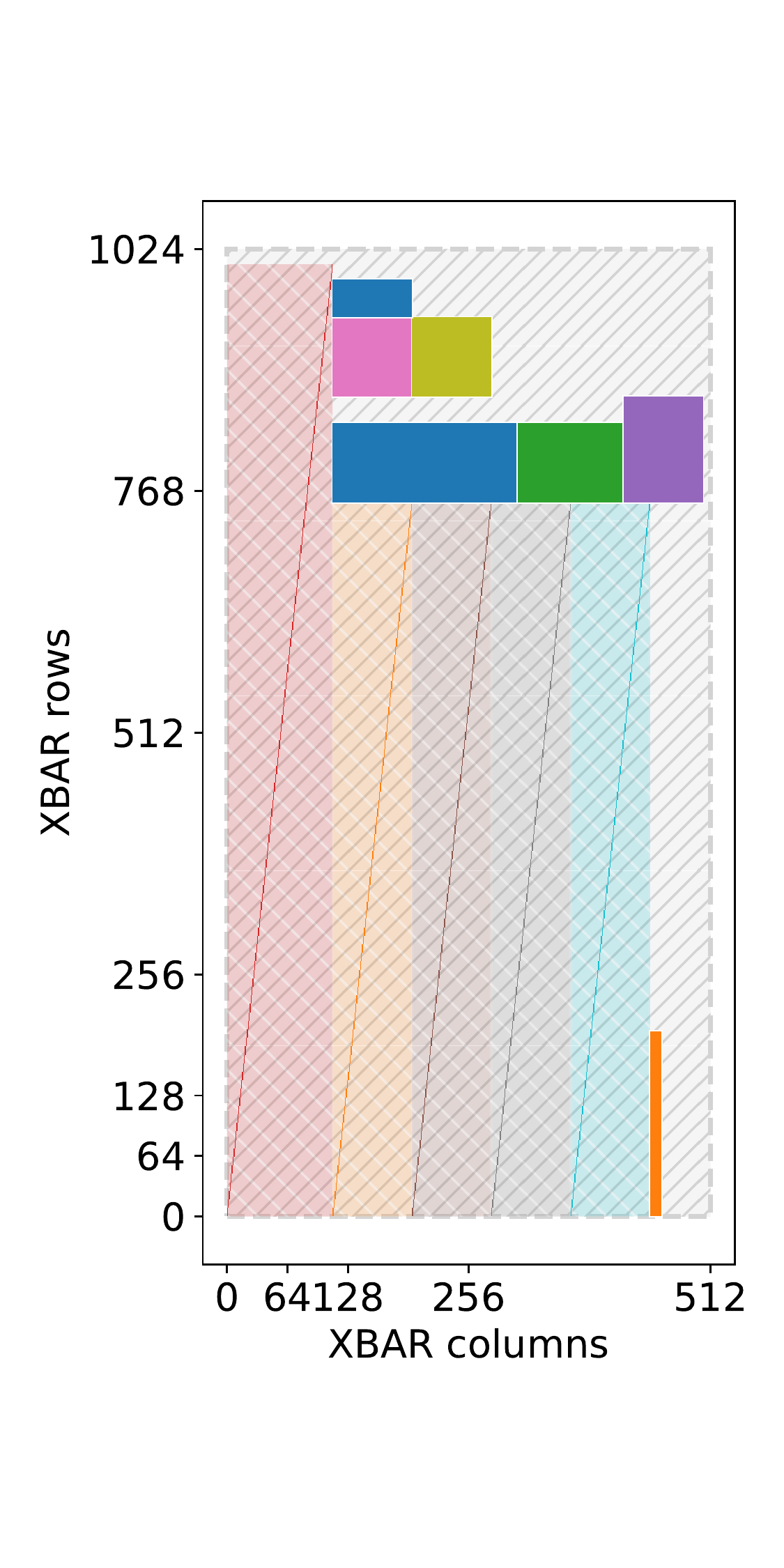}}
    \end{minipage}%
    \begin{minipage}{1.2\columnwidth}
    \centering
    \subfloat[MicroNet-KWS-S deployment on multiple $128 \times 128$ NVM crossbar  tiles.]{\label{fig:depth_wise_kws:b}\includegraphics[width=0.98\columnwidth]{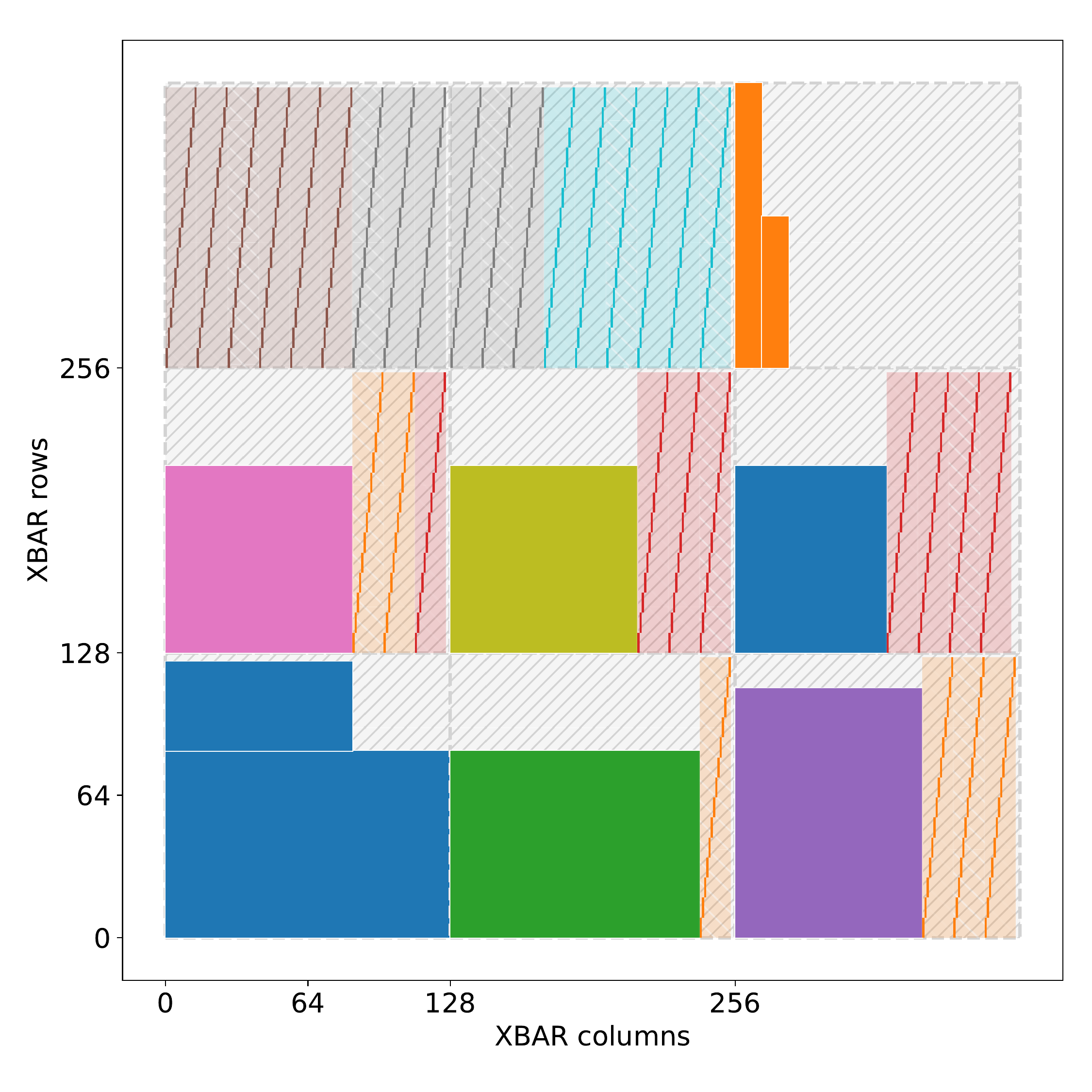}}
    \qquad
    \centering
    \subfloat[MicroNet-KWS-S deployment on multiple $64 \times 64$ NVM crossbar  tiles.]{\label{fig:depth_wise_kws:c}\includegraphics[width=0.98\columnwidth]{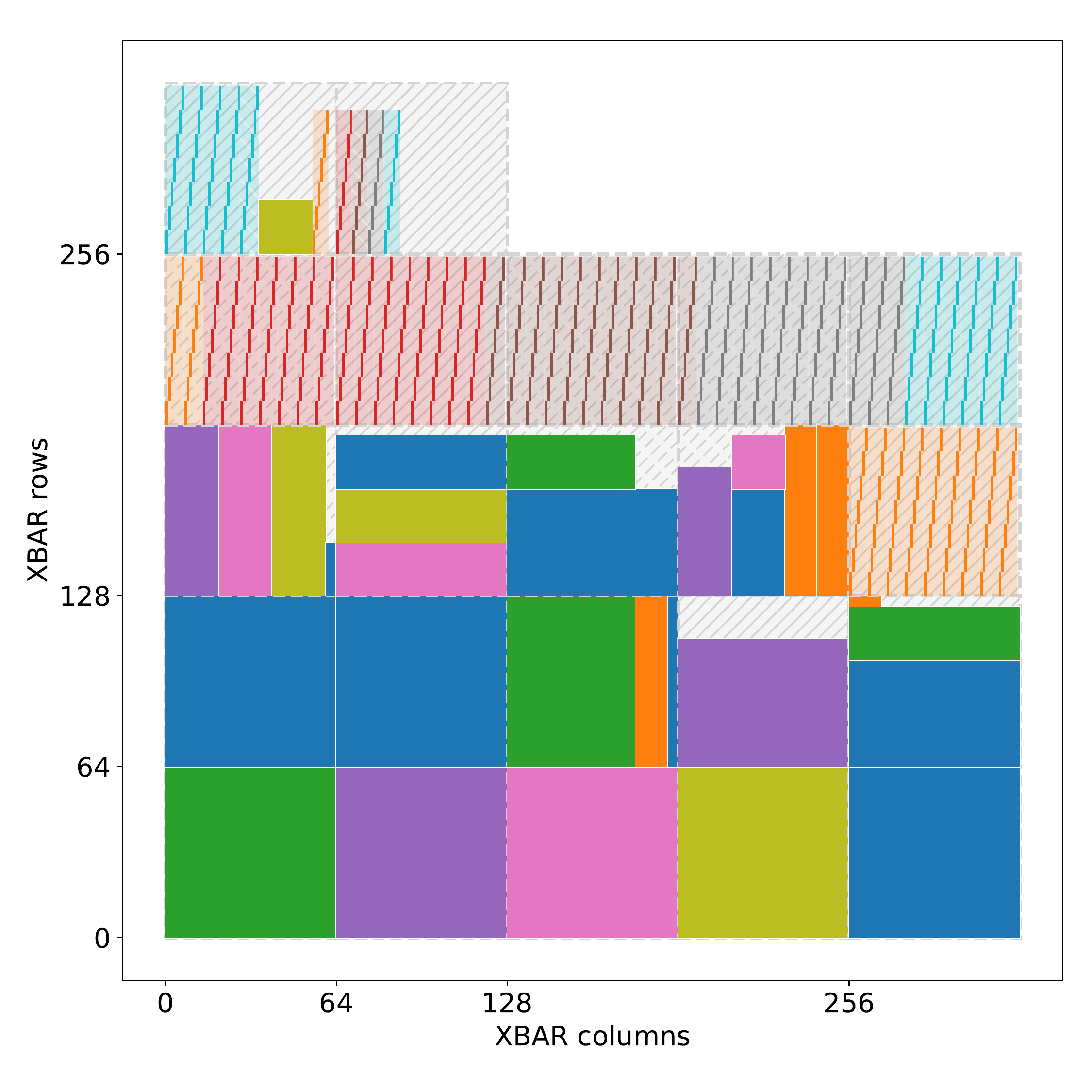}}
    \end{minipage}\par\medskip
	
    \caption{MicroNet-KWS-S deployment on a $1024 \times 512$ differential crossbar array. Depthwise convolutional layers are represented using semi-transparent colors, while their effective weights --non-zero weights contributing to the layer computation-- are depicted as solid diagonal areas. Layers represented with solid colors correspond to regular 2-dimensional layers. }
    \label{fig:depth_wise_kws}
\end{figure*}

As described in Section~\ref{sec:models}, depthwise  convolutional layers (DWCL)  are  not  suitable  for  CiM analog architectures, attending both at effective utilization and SNR contributions. 
This effect is clearly visible in Figure~\ref{fig:depth_wise_kws:a}, which magnifies how DWCLs are inefficient compared against normal convolutional layers. The weights deployment map describes how \emph{MicroNet-KWS-S} NN reaches an effective utilization for MicroNet-KWS-S as low as $~9\%$.
A possible solution to better use the NVM crossbar area resources require splitting the large DWCL GEMM operations into smaller ones operating in a sequential way. Figures~\ref{fig:depth_wise_kws:b} and ~\ref{fig:depth_wise_kws:c} manifest how the crossbar area utilization improves inversely with the maximum size of the split GEMM operation. The main drawback associated with this technique is the layer latency, which gets magnified due to the sequential operation scheme.
Table \ref{table:dwcl} summarizes the utilization/latency increment trade-off.
\begin{table}[ht]
\caption{Resource vs latency comparison for the deployment of MicroNet-KWS-S as an example of NN heavily relying on Depthwise convolutional layers.}
\centering
\begin{tabular}{|c|c|c|c|}
\hline
\textbf{Crossbar}               & \textbf{$1024 \times 512$} & $128 \times 128$ & $64 \times 64$ \\
\hline
\textbf{Eff. Utilization}    & $9\%$            & $40\%$          & $66 \%$ \\
\hline
\textbf{Inference/s}    & $4122$            & $1467$          & $642$ \\
\hline
\end{tabular}
\label{table:dwcl}
\end{table}

%%%%%%%%%%%%%%%%%%%%%%%%%%%%%%%%%%%%%%%%%%%%%%%%%%%%%%%%%%%%%%%%%%%%%%%%%%%%%%%
%%%%%%%%%%%%%%%%%%%%%%%%%%%%%%%%%%%%%%%%%%%%%%%%%%%%%%%%%%%%%%%%%%%%%%%%%%%%%%%

\end{document}